\RequirePackage{fix-cm} 
\documentclass[a4paper, twoside, reqno, 12pt, dvips]{amsart}
\usepackage{fixltx2e}     

\usepackage[english]{babel}
\usepackage[latin1]{inputenc}

\usepackage{eucal}
\usepackage{esint}
\usepackage{amsgen}
\usepackage{amsthm}
\usepackage{amssymb}
\usepackage{amsmath}
\usepackage{amsfonts}
\usepackage{verbatim}
\usepackage{mathrsfs, dsfont}

\usepackage{a4wide}
\usepackage{xspace}

\usepackage{indentfirst}
\usepackage{graphicx}
\usepackage{subfigure}
\usepackage[section]{placeins}
\usepackage{psfrag}

\usepackage{microtype}
\usepackage[bf, up, hang]{caption}

\usepackage[usenames, dvipsnames, pdf]{pstricks}
\usepackage{epsfig}
\usepackage{pst-grad} 
\usepackage{pst-plot} 

\usepackage{color}
\definecolor{oneblue}{rgb}{0.0, 0.0, 0.85}
\definecolor{darkgrey}{rgb}{0.273, 0.281, 0.30}

\usepackage[colorlinks,
            urlcolor=oneblue,
            linkcolor=oneblue,
            citecolor=oneblue,
            bookmarksopen=false,
            pagebackref]{hyperref}

\usepackage[compact]{titlesec}

\titleformat{\section}{\normalfont\Large\bfseries\sffamily\center\color{darkgrey}}{\thesection.}{0.5em}{}{}
\titleformat{\subsection}{\normalfont\large\bfseries\sffamily\color{darkgrey}}{\thesubsection.}{0.4em}{}{}
\titleformat{\subsubsection}{\normalfont\normalsize\bfseries\sffamily\color{darkgrey}}{\thesubsubsection.}{0.3em}{}{}

\titlespacing*{\section}{1.0em}{1.0em}{0.8em}[0em]
\titlespacing*{\subsection}{1.0em}{1.0em}{0.8em}[0em]
\titlespacing*{\subsubsection}{1.0em}{0.7em}{0.6em}[0em]

\usepackage{titletoc}

\contentsmargin{0.0em}
\dottedcontents{section}[2.5em]{\addvspace{0.45em}\bfseries}{1.0em}{0pc}
\dottedcontents{subsection}[4.5em]{}{2.6em}{0pc}
\dottedcontents{subsubsection}[5.5em]{}{3.0em}{0pc}

\usepackage{fancyhdr}
\usepackage{lastpage}

\newcommand*\Title{Numerical schemes for the Serre equations}
\newcommand*\Authors{D.~Dutykh, D.~Clamond, P.~Milewski \& D.~Mitsotakis}

\usepackage{acronym}

\numberwithin{equation}{section}

\newtheorem{remark}{Remark}

\newcommand{\up}[1]{$\,^{\mathrm{\small\textsf{#1}}}$} 

\newcommand{\us}{\tilde{u}}
\newcommand{\hm}{\bar{\/h}}
\newcommand{\um}{\bar{\/u}}
\newcommand{\qm}{\bar{\/q}}
\newcommand{\vs}{\tilde{v}}
\newcommand{\Z}{\mathds{Z}}
\newcommand{\R}{\mathds{R}}

\newcommand{\A}{\mathbb{A}}
\newcommand{\M}{\mathbb{M}}
\newcommand{\D}{\mathbb{D}}
\newcommand{\I}{\mathbb{I}}
\newcommand{\ui}{\mathrm{i}}
\newcommand{\C}{\mathcal{C}}
\newcommand{\ud}{\mathrm{d}}
\newcommand{\ue}{\mathrm{e}}
\newcommand{\F}{\mathcal{F}}
\newcommand{\Q}{\mathcal{Q}}

\newcommand{\N}{\mathcal{N}}

\newcommand{\Mr}{\mathcal{M}}
\newcommand{\Fr}{\mathcal{F}}
\renewcommand{\u}{\mathbf{u}}

\renewcommand{\H}{\mathcal{H}}
\renewcommand{\L}{\mathcal{L}}
\newcommand{\phim}{\bar{\phi}}
\newcommand{\eps}{\varepsilon}
\renewcommand{\O}{\mathcal{O}}

\newcommand{\nus}{\tilde{\nu}}
\newcommand{\nub}{\check{\nu}}
\renewcommand{\Pr}{\ensuremath{\mathcal{P}}}
\newcommand{\x}{\boldsymbol{x}}
\newcommand{\vu}{\boldsymbol{u}}
\newcommand{\phib}{\check{\phi}}
\newcommand{\phis}{\tilde{\phi}}
\newcommand{\vmu}{\boldsymbol{\mu}}
\newcommand{\vum}{\bar{\boldsymbol{u}}}

\newcommand{\vmum}{\bar{\boldsymbol{\mu}}}
\newcommand{\vmus}{\tilde{\boldsymbol{\mu}}}
\newcommand{\vmub}{\check{\boldsymbol{\mu}}}

\newcommand{\dt}{\partial_t}
\newcommand{\dy}{\partial_y}
\renewcommand{\div}{\grad\scal}
\newcommand{\sech}{\mathrm{sech}}
\newcommand{\diag}{\mathrm{diag}}
\newcommand{\scal}{\boldsymbol{\cdot}}
\newcommand{\grad}{\boldsymbol{\nabla}}
\newcommand{\sign}{\mathop{\mathrm{sign}}}

\newcommand{\minmod}{\mathop{\mathrm{minmod}}}
\newcommand{\pd}[2]{\frac{\partial\, #1}{\partial\/ #2}}
\newcommand{\od}[2]{\frac{\mathrm{d}\,#1}{\mathrm{d}\/#2}}

\newcommand{\half}{{\textstyle{1\over2}}}
\newcommand{\frth}{{\textstyle{3\over4}}}
\newcommand{\third}{{\textstyle{1\over3}}}
\newcommand{\sixth}{{\textstyle{1\over6}}}

\begin{document}

\title[\Title]{Finite volume and pseudo-spectral schemes for the fully nonlinear 1D Serre equations}

\author[D.~Dutykh]{Denys Dutykh$^*$}
\address{University College Dublin, School of Mathematical Sciences, Belfield, Dublin 4, Ireland \and LAMA, UMR 5127 CNRS, Universit\'e de Savoie, Campus Scientifique, 73376 Le Bourget-du-Lac Cedex, France}
\email{Denys.Dutykh@ucd.ie}
\urladdr{http://www.denys-dutykh.com/}
\thanks{$^*$ Corresponding author}

\author[D.~Clamond]{Didier Clamond}
\address{Laboratoire J.-A. Dieudonn\'e, Universit\'e de Nice -- Sophia Antipolis, Parc Valrose, 06108 Nice cedex 2, France}
\email{diderc@unice.fr}
\urladdr{http://math.unice.fr/~didierc/}

\author[P.~Milewski]{Paul Milewski}
\address{Deptartment of Mathematical Sciences, University of Bath, Bath, BA2 7JX, UK}
\email{P.A.Milewski@bath.ac.uk}

\author[D.~Mitsotakis]{Dimitrios Mitsotakis}
\address{University of California, Merced, 5200 North Lake Road, Merced, CA 94353, USA}
\email{dmitsot@gmail.com}
\urladdr{http://dmitsot.googlepages.com/}

\begin{abstract}
After we derive the Serre system of equations of water wave theory from a generalized variational principle, we present some of its structural properties. We also propose a robust and accurate finite volume scheme to solve these equations in one horizontal dimension. The numerical discretization is validated by comparisons with analytical, experimental data or other numerical solutions obtained by a highly accurate pseudo-spectral method.

\bigskip
\noindent \textbf{\keywordsname:} Serre equations; finite volumes; UNO scheme; IMEX scheme; spectral methods; Euler equations; free surface flows
\end{abstract}


\maketitle
\tableofcontents
\thispagestyle{empty}

\section{Introduction}

The full water wave problem consisting of the Euler equations with a free surface still is a very difficult to study theoretically and even numerically. Consequently, water wave theory has always been developed through the derivation, analysis and comprehension of various approximate models (see the historical review of \textsc{Craik} \cite{Craik2004} for more information). For this reason, a plethora of approximate models have been derived under various physical assumptions. In this family, the Serre equations have a particular place and they are the subject of the present study. The Serre equations can be derived from the Euler equations, contrary to Boussinesq systems or the shallow water system, without the small amplitude or the hydrostatic assumptions respectively.

The Serre equations are named after Fran\c{c}ois \textsc{Serre}, an engineer at \'{E}cole Nationale des Ponts et Chauss\'ees, who derived this model for the first time in 1953 in his prominent paper entitled {\em ``Contribution \`a l'\'etude des \'ecoulements permanents et variables dans les canaux''} (see \cite{Serre1953}). Later, these equations were independently rediscovered by \textsc{Su} and \textsc{Gardner} \cite{Su1969} and by \textsc{Green}, \textsc{Laws} and \textsc{Naghdi} \cite{Green1974}. The extension of Serre equations for general uneven bathymetries was derived by \textsc{Seabra-Santos} \emph{et al.} \cite{Seabra-Santos1987}. In the Soviet literature these equations were known as the Zheleznyak-Pelinovsky model \cite{Zheleznyak1985}. For some generalizations and new results we refer to recent studies by \textsc{Barth\'el\'emy} \cite{Barthelemy2004}, \textsc{Dias} \& \textsc{Milewski} \cite{Dias2010} and \textsc{Carter} \& \textsc{Cienfuegos} \cite{Carter2011}.

A variety of numerical methods have been applied to discretize dispersive wave models and, more specifically, the Serre equations. A pseudo-spectral method was applied in \cite{Dias2010}, an implicit finite difference scheme in \cite{Mirie1982, Barthelemy2004} and a compact higher-order scheme in \cite{CBB1, CBB2}. Some Galerkin and Finite Element type methods have been successfully applied to Boussinesq-type equations \cite{DMII, Mitsotakis2007, Avilez-Valente2009, ADM2}. A finite difference discretization based on an integral formulation was proposed by \textsc{Bona} \& \textsc{Chen} \cite{BC}.

Recently, efficient high-order explicit or implicit-explicit finite volume schemes for dispersive wave equations have been developed \cite{ChazelLannes2010, Dutykh2010e, Dutykh2010e}. The robustness of the proposed numerical schemes also allowed simulating the run-up of long waves on a beach with high accuracy \cite{Dutykh2010e}. The present study is a further extension of the finite volume method to the practically important case of the Serre equations. We develop also a pseudo-spectral Fourier-type method to validate the proposed finite volume scheme. In all cases where the spectral method is applicable, it outperforms the finite volumes. However, the former is applicable only to smooth solutions in periodic domains, while the area of applicability of the latter is much broader including dispersive shocks (or undular bores) \cite{El2006}, non-periodic domains, etc.

The present paper is organized as follows. In Section \ref{sec:model} we provide a  derivation of the Serre equations from a relaxed Lagrangian principle and discuss some structural properties of the governing equations. The rationale on the employed finite volume scheme are given in Section \ref{sec:nums}. A very accurate pseudo-spectral method for the numerical solution of the Serre equations is presented in Section \ref{sec:spectral}. In Section \ref{sec:numres}, we present convergence tests and numerical experiments validating the model and the numerical schemes. Finally, Section \ref{sec:concl} contains the main conclusions.

\section{Mathematical model}\label{sec:model}

\begin{figure}
\begin{center}
\scalebox{1} 
{
\begin{pspicture}(0,-2.914)(13.719063,2.9334376)
\definecolor{color245g}{rgb}{0.5568627450980392,0.48627450980392156,0.43529411764705883}
\definecolor{color245f}{rgb}{0.2,0.12549019607843137,0.050980392156862744}
\definecolor{color255}{rgb}{0.027450980392156862,0.08627450980392157,0.6901960784313725}
\psline[linewidth=0.027999999cm,linestyle=dashed,dash=0.16cm 0.16cm,arrowsize=0.05291667cm 2.0,arrowlength=1.4,arrowinset=0.4]{->}(0.0,1.38)(13.52,1.4)
\psline[linewidth=0.027999999cm,linestyle=dashed,dash=0.16cm 0.16cm,arrowsize=0.05291667cm 2.0,arrowlength=1.4,arrowinset=0.4]{<-}(6.58,2.9)(6.54,-2.9)
\psframe[linewidth=0.024,dimen=outer,fillstyle=gradient,gradlines=2000,gradbegin=color245g,gradend=color245f,gradmidpoint=1.0](13.28,-2.2)(0.06,-2.62)
\usefont{T1}{ppl}{m}{n}
\rput(13.284532,0.97){$\x$}
\usefont{T1}{ppl}{m}{n}
\rput(6.224531,2.73){$y$}
\usefont{T1}{ppl}{m}{n}
\rput(6.2145314,1.19){$O$}
\pscustom[linewidth=0.07,linecolor=color255]
{
\newpath
\moveto(0.0,1.02)
\lineto(0.18,1.12)
\curveto(0.27,1.17)(0.455,1.265)(0.55,1.31)
\curveto(0.645,1.355)(0.825,1.475)(0.91,1.55)
\curveto(0.995,1.625)(1.245,1.715)(1.41,1.73)
\curveto(1.575,1.745)(1.9,1.73)(2.06,1.7)
\curveto(2.22,1.67)(2.545,1.545)(2.71,1.45)
\curveto(2.875,1.355)(3.175,1.21)(3.31,1.16)
\curveto(3.445,1.11)(3.7,1.05)(3.82,1.04)
\curveto(3.94,1.03)(4.175,1.035)(4.29,1.05)
\curveto(4.405,1.065)(4.72,1.15)(4.92,1.22)
\curveto(5.12,1.29)(5.585,1.42)(5.85,1.48)
\curveto(6.115,1.54)(6.585,1.625)(6.79,1.65)
\curveto(6.995,1.675)(7.355,1.66)(7.51,1.62)
\curveto(7.665,1.58)(8.01,1.465)(8.2,1.39)
\curveto(8.39,1.315)(8.97,1.2)(9.36,1.16)
\curveto(9.75,1.12)(10.325,1.135)(10.51,1.19)
\curveto(10.695,1.245)(11.09,1.425)(11.3,1.55)
\curveto(11.51,1.675)(11.9,1.845)(12.08,1.89)
\curveto(12.26,1.935)(12.61,1.99)(12.78,2.0)
\curveto(12.95,2.01)(13.165,2.025)(13.3,2.04)
}
\psline[linewidth=0.027999999cm,arrowsize=0.05291667cm 2.0,arrowlength=1.4,arrowinset=0.4]{<->}(3.88,1.34)(3.84,-2.16)
\usefont{T1}{ppl}{m}{n}
\rput(3.5445313,-0.53){$d$}
\psline[linewidth=0.027999999cm,arrowsize=0.05291667cm 2.0,arrowlength=1.4,arrowinset=0.4]{<->}(12.3,1.88)(12.28,-2.16)
\usefont{T1}{ppl}{m}{n}
\rput(11.624531,-0.29){$h(\x,t)$}
\psline[linewidth=0.027999999cm,arrowsize=0.05291667cm 2.0,arrowlength=1.4,arrowinset=0.4]{<->}(1.6,1.36)(1.6,1.72)
\psline[linewidth=0.027999999cm](1.6,1.76)(1.6,2.16)
\usefont{T1}{ppl}{m}{n}
\rput(0.9945313,2.01){$\eta(\x,t)$}
\end{pspicture} 
}
\end{center}
\caption{\em Sketch of the physical domain.}
\label{fig:sketch}
\end{figure}
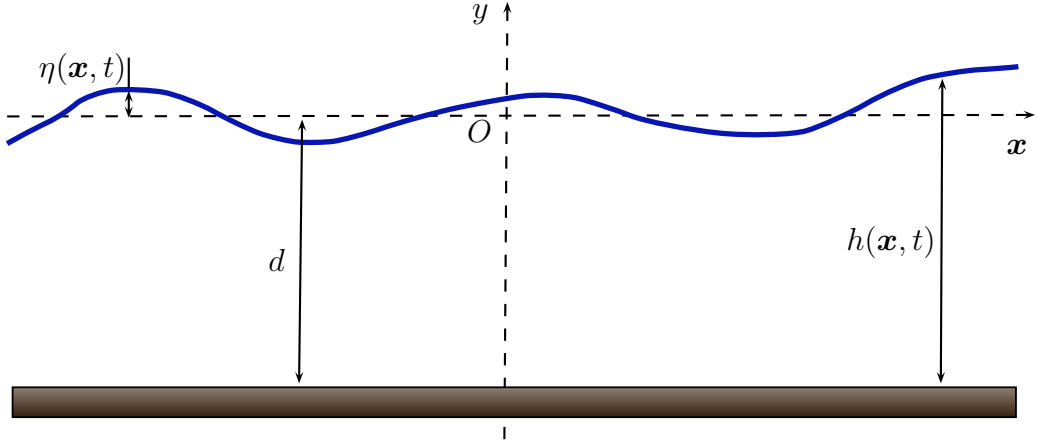

Consider an ideal incompressible fluid of constant density $\rho$. The vertical projection of the fluid domain $\Omega$ is a subset of $\R^2$. The horizontal independent variables are denoted by $\x = (x_1,x_2)$ and the upward vertical one by $y$. The origin of the Cartesian coordinate system is chosen such that the surface $y=0$ corresponds to the still water level. The fluid is bounded below by an impermeable bottom at $y = -d$ and above by the free surface located at $y = \eta (\x,t)$. We assume that the total depth $h(\x, t) \equiv d + \eta (\x,t)$ remains positive $h (\x,t) \geqslant h_0>0$ at all times $t$. The sketch of the physical domain is shown in Figure~\ref{fig:sketch}.

\begin{remark}
We make the classical assumption that the free surface is a graph $y = \eta (\x,t)$ of a single-valued function. This means in practice that we exclude some interesting phenomena, (e.g., wave breaking) which are out of the scope of this modeling paradigm.
\end{remark}

Assuming that the flow is incompressible and irrotational, the governing equations of the classical water wave problem are the following \cite{Lamb1932, Stoker1958, Mei1994, Whitham1999}
\begin{align}
  \grad^2\phi\ +\ \dy^{\,2}\,\phi\ &=\ 0 \qquad
  -d(\x, t)\,\leqslant y\leqslant\eta(\x,t), \label{eq:laplace} \\
  \dt\eta\ +\ (\grad\phi)\scal(\grad\eta)\ -\ \dy\,\phi\ &=\ 0 
  \qquad y = \eta(\x, t), \label{eq:kinematic} \\
  \dt\phi\ +\ \half\,|\grad\phi|^2\ +\ \half\,(\dy\phi)^2\ +\ g\,\eta\ &=\ 0 
  \qquad y = \eta(\x,t), \label{eq:bernoulli} \\
  d_t\ +\ (\grad d)\scal(\grad\phi)\ +\ \dy\,\phi\ &=\ 0 
  \qquad y = -d(\x, t), \label{eq:bottomkin}
\end{align}
with $\phi$ being the velocity potential (by definition, the irrotational velocity field $(\u, v) = (\grad\phi, \dy\phi)$, $g$ the acceleration due to the gravity force and $\grad = (\partial_{x_1}, \partial_{x_2})$ denotes the gradient operator in horizontal Cartesian coordinates and $|\grad\phi|^2 \equiv (\grad\phi)\cdot(\grad\phi)$.

The incompressibility condition leads to the Laplace equation for $\phi$. The main difficulty of the water wave problem lies on the nonlinear free surface boundary conditions and that the free surface shape is unknown. Equations \eqref{eq:kinematic} and \eqref{eq:bottomkin} express the free-surface kinematic condition and bottom impermeability respectively, while the dynamic condition \eqref{eq:bernoulli} expresses the free surface isobarity.

The water wave problem possesses several variational structures \cite{Petrov1964, Whitham1965, Luke1967, Zakharov1968, Broer1974}. In the present study, we will focus mainly on the Lagrangian variational formalism but not exclusively. The surface gravity wave equations \eqref{eq:laplace}--\eqref{eq:bottomkin} can be derived by minimizing the following functional proposed by Luke \cite{Luke1967}:
\begin{equation}\label{eq:L0}
\mathcal{L}\,=\int_{t_1}^{t_2}\!\int_{\Omega}\mathscr{L}\,\rho\ \ud^2\/\x\,\ud\/t, \quad
\mathscr{L}\,=\,-\int_{-d}^\eta\left[\,g\,y\, +\, \dt\,\phi\, +\, 
\half\,(\grad\,\phi)^2\, +\, \half\,(\dy\,\phi)^2\,\right]\,\ud\/y.
\end{equation}
In a recent study, Clamond and Dutykh \cite{Clamond2009} proposed using Luke's Lagrangian \eqref{eq:L0} in the following relaxed form
\begin{align}\label{eq:L3}
\mathscr{L}\ =&\ (\eta_t+\vmus\scal\grad\eta-\nus)\,\phis\ +\ (d_t+\vmub\scal\grad d+\nub)\,\phib\ -\ 
\half\,g\,\eta^2\  \nonumber \\ 
& +\ \int_{-d}^{\,\eta} \left[\,\vmu\scal\vu-{\half}\/\vu^2\,+\,\nu\/v-\half\/v^2\, +
\,(\div\vmu+\nu_y)\,\phi\,\right]\ud\/y,
\end{align}
where \{$\vu, v, \vmu, \nu$\} are the horizontal, vertical velocities and associated Lagrange multipliers, respectively. The additional variables \{$\vmu, \nu$\} (Lagrange multipliers) are called pseudo-velocities. The `tildes' and `wedges' denote, respectively, a quantity computed at the free surface $y = \eta(\x, t)$ and at the bottom $y = -d(\x, t)$. We shall also denote below with `bars' the quantities averaged over the water depth.

While the original Lagrangian \eqref{eq:L0} incorporates only two variables ($\eta$ and $\phi$), the relaxed Lagrangian density \eqref{eq:L3} involves six variables \{$\eta, \phi, \vu, v, \vmu, \nu$\}. These additional degrees of freedom provide us with more flexibility in constructing various approximations. For more details, explanations and examples we refer to \cite{Clamond2009}.

\subsection{Derivation of the Serre equations}

Now, we illustrate the practical use of the variational principle \eqref{eq:L3} on an example borrowed from \cite{Clamond2009}. First of all, we choose a simple shallow water ansatz, which is a zeroth-order polynomial in $y$ for $\phi$ and for $\vu$, and a first-order one for $v$, i.e., we approximate flows that are nearly uniform along the vertical direction
\begin{equation}\label{eq:anssha}
\phi\ \approx\ \phim(\x,t), \qquad \vu\ \approx\ \vum(\x,t), \qquad v\ \approx\ (y+d)\,(\eta+d)^{-1}\ \vs(\x,t).
\end{equation}
We have also to introduce suitable ansatz for the Lagrange multiplier $\vmu$ and $\nu$
\begin{equation*}
\vmu\ \approx\ \vmum(\x,t), \qquad \nu\ \approx\ (y+d)\,(\eta+d)^{-1}\ \nus(\x,t).
\end{equation*}
In the remainder of this paper, we will assume for simplicity the bottom to be flat $d(\x,t) = d = \text{Cst}$ (the application of this method to uneven bottoms can be found in \cite{Dutykh2011b, Dutykh2013}, for example). With this ansatz the Lagrangian density \eqref{eq:L3} becomes
\begin{eqnarray}\label{eq:L3sw}
\mathscr{L}\ &=&\ (\eta_t+\vmum\scal\grad\eta)\,\phim\ -\ \half\,g\,\eta^2\  \nonumber \\
&&+\ (\eta+d)
\left[\,\vmum\scal\vum\ -\ {\half}\,\vum^2\ +\ \third\,\nus\,\vs\ -\ \sixth\,\vs^2\, +\ 
\phim\,\grad\scal\vmum\,\right].
\end{eqnarray}
Finally, we impose a constraint of the free surface impermeability, i.e.
\begin{equation*}
\nus\ =\ \eta_t\ +\ \vmum\scal\grad\eta.
\end{equation*}
After substituting the last relation into the Lagrangian density \eqref{eq:L3sw}, the Euler--Lagrange equations and some algebra lead to the following equations:
\begin{eqnarray} \label{eq:swser}
h_t\ +\ \grad\scal[\,h\,\vum\,]\ &=&\ 0,  \\
\vum_t\ +\ \half\,\grad|\vum|^2\ +\ g\,\grad h\ +\
\third\,h^{-1}\,\grad[\,h^2\,\tilde{\gamma}\,]\ &=&\ (\vum\scal\grad h)\,\grad(h\grad\scal\vum) \nonumber\\
&&\ -\ \left[\,\vum\scal\grad(h\grad\scal\vum)\,\right]\grad h,\label{eq:swser2}
\end{eqnarray}
where we eliminated $\phim$, $\vmum$ and $\vs$ and where
\begin{equation}\label{eq:gamma}
\tilde{\gamma}\ \equiv\ \vs_t\ +\ \vum\scal\grad\vs \ =\ h\ \left\{\
(\grad\scal\vum)^2\,-\,\grad\scal\vum_t\,-\,\vum\scal\grad\,[\,\grad\scal\vum\,]\ \right\},
\end{equation}
is the fluid vertical acceleration at the free surface. The vertical velocity at the free surface $\vs$ can be expressed in terms of other variables as well, i.e.,
\begin{equation*}
  \vs\ =\ \frac{\eta_t\, +\, (\grad\phim)\scal(\grad\eta)}{1\, + \,\third|\grad\eta|^2}.
\end{equation*}

In two dimensions (one horizontal dimension), the sum of two terms on the right hand side of \eqref{eq:swser2} vanish and the system \eqref{eq:swser}, \eqref{eq:swser2} reduces to the classical Serre equations \cite{Serre1953}.

\begin{remark}
In \cite{Clamond2009} it is explained why equations \eqref{eq:swser}, \eqref{eq:swser2} cannot be obtained from the classical Luke's Lagrangian. One of the main reasons is that the horizontal velocity $\vum$ does not derive from the potential $\phim$ using a simple gradient operation. Thus, a relaxed form of the Lagrangian density \eqref{eq:L3} is necessary for the variational derivation of the Serre equations \eqref{eq:swser}, \eqref{eq:swser2} (see also \cite{Kim2001} \& \cite{Miles1985}).
\end{remark}

\begin{remark}
In some applications in coastal engineering it is required to estimate the loading exerted by water waves onto vertical structures \cite{Clauss2011}. The pressure can be computed in the framework of the Serre equations as well. For the first time these quantities were computed in the pioneering paper by M.~\textsc{Zheleznyak} (1985) \cite{Zheleznyak1985a}. Here for simplicity we provide the expressions in two space dimensions which were derived in \cite{Zheleznyak1985a}. The pressure distribution inside the fluid column being given by
\begin{equation*}
  \frac{\Pr(x,y,t)}{\rho\/g\/d}\ =\ \frac{\eta-y}{d}\ +\ 
  \frac{1}{2}\left[\left(\frac{h}{d}\right)^{\!2}\ -\ 
  \left(1 + \frac{y}{d}\right)^{\!2}\,\right]\,\frac{\tilde{\gamma}\,d}{g\,h},
\end{equation*}
one can compute the force $\Fr$ exerted on a vertical wall:
\begin{equation*}
\frac{\Fr(x,t)}{\rho\/g\/d^{\/2}}\ =\,\int_{-d}^{\eta}\frac{\Pr}{\rho\/g\/d^{\/2}}\,\ud y\ = \,\left(\frac{1}{2}+\frac{\tilde{\gamma}}{3\,g}\right)\!\left(\frac{h}{d}\right)^{\!2}.
\end{equation*}
Finally, the tilting moment $\Mr$ relative to the sea bed is given by the following formula:
\begin{equation*}
\frac{\Mr(x,t)}{\rho\/g\/d^{\/3}}\ =\,\int_{-d}^{\eta}\frac{\Pr}{\rho\/g\/d^{\/3}}\,(y+d)\,\ud y
\ = \,\left(\frac{1}{6}+\frac{\tilde{\gamma}}{8\,g}\right)\!\left(\frac{h}{d}\right)^{\!3}.
\end{equation*}
\end{remark}

\subsubsection{Generalized Serre equations}

A further generalization of the Serre equations can be obtained if we modify slightly the shallow water ansatz \eqref{eq:anssha} following again the ideas from \cite{Clamond2009}:
\begin{align*}
\phi\ \approx\ \phim(x,t), \qquad
u\ \approx\ \um(x,t), \qquad
v\ \approx\,\left[\,\frac{y+d}{\eta+d}\,\right]^\lambda\, \vs(x,t).
\end{align*}
In the following we consider for simplicity the two-dimensional case and put $\mu=u$ and $\nu=v$ together with the constraint $\vs = \eta_t + \us\/\eta_x$ (free surface impermeability). Thus, the Lagrangian density \eqref{eq:L3} becomes
\begin{equation}\label{lagdensermod}
\mathscr{L}\ =\ \left(\,h_t\,+\,[\,h\,\um\,]_x\/\right)\phis\, -\, \half\,g\,\eta^2\, 
+\, \half\,h\,\um^2\, +\, \half\,\beta\,h\left(\,\eta_t\,+\,\um\,\eta_x\,\right)^2,
\end{equation}
where $\beta=(2\lambda+1)^{-1}$. After some algebra, the Euler--Lagrange equations lead to the following equations
\begin{eqnarray}\label{eq:sermod}
h_t\ +\ [\,h\,\um\,]_x\, &=\ 0, \\
\um_t\ +\ \um\,\um_x\ +\ g\,h_x\ +\ \beta\,h^{-1}\,[\,h^2\,\tilde{\gamma}\,]_x\, &=\ 0, \label{eq:sermod2}
\end{eqnarray}
where $\tilde{\gamma}$ is defined as above \eqref{eq:gamma}. If $\beta=\third$ (or, equivalently, $\lambda=1$) the classical Serre equations \eqref{eq:swser}, \eqref{eq:swser2} are recovered.

Using equations \eqref{eq:sermod} and \eqref{eq:sermod2} one can show that the following relations hold
\begin{equation*}
  [\,h\,\um\,]_t\ +\ \left[\,h\,\um^2\, +\, \half\,g\,h^2\, +\, 
  \beta\, h^2\,\tilde{\gamma}\,\right]_x\ =\ 0,
\end{equation*}
\begin{equation*}
  [\,\um\, -\, \beta\, h^{-1}(h^3\um_x)_x\,]_t\ +\ \left[\,\half\,\um^2\, +\, g\,h\, -\, 
  \half\, h^2\,\um_x^2\, -\, \beta\,\um h^{-1}\,(h^3\um_x)_x\,\right]_x\ =\ 0,
\end{equation*}
\begin{equation}\label{eq:vflux}
  [\,h\,\um\, -\, \beta\,(h^3\um_x)_x\,]_t\ +\ \left[\,h\,\um^2\, +\, \half\,g\,h^2\, -\, 
  2\,\beta\,h^3\,\um_x^2\, -\, \beta\, h^3\,\um\,\um_{xx}\, -\, h^2\,h_x\,\um\,\um_x\,\right]_x\ =\ 0,
\end{equation}
\begin{equation*}
  \left[\,\half\,h\,\um^2\, +\, \half\,\beta\, h^3\,\um_x^2\, +\, \half\, g\,h^2\,\right]_t\ +\ 
  \left[\,\left(\,\half\,\um^2\, +\, \half\,\beta\, h^2\,\um_x^2\, +\, g\,h\, +\, 
  \beta\, h\,\tilde{\gamma}\,\right)\,h\,\um\,\right]_x\ =\ 0.
\end{equation*}
Physically, these relations represent conservations of the momentum, quantity $\qm = \um\, - \,\beta\,h^{-1}(h^3\um_x)_x$, its flux $\tilde{q} := h\,\um\, - \,\beta\,(h^3\um_x)_x$ and the total energy, respectively. Moreover, the Serre equations are invariant under the Galilean transformation. This property is naturally inherited from the full water wave problem since our ansatz does not destroy this symmetry \cite{Benjamin1982} and the derivation is made according to variational principles.

Equations \eqref{eq:sermod}--\eqref{eq:sermod2} admit a ($2\pi/k$)-periodic cnoidal traveling wave solution
\begin{align}\label{solsermod}
\um\ &=\ \frac{c\,\eta} {d+\eta},\\
\eta\ &=\ a\, \frac{\operatorname{dn}^2\! \left(\/\half\/\varkappa\/(x-ct)\/|\/m\/\right) - E/K} {1-E/K}\ =\ a\, -\, H\,\operatorname{sn}^2\! \left(\/\half\/\varkappa\/(x-ct)\/|\/m\/\right), \label{solsermod2}
\end{align}
where $\text{dn}$ and $\text{sn}$ are the Jacobian elliptic functions with parameter $m$ ($0\leqslant m\leqslant1$), and where $K = \text{K}\/(m)$ and $E = \text{E}\/(m)$ are the complete elliptic integrals of the first and second kind, respectively \cite{Abramowitz1965}. The wave parameters are given by the relations
\begin{eqnarray}\label{relparasolsermod}
k\ =\ \frac{\pi\,\varkappa}{2\,K}, \qquad  H\ =\ \frac{m\,a\,K}{K-E}, \qquad
(\varkappa\/d)^2\ =\ \frac{g\,H}{m\,\beta\,c^2},\\
\gdef\thesubequation{\theequation\textit{d}} m\ =\ \frac{g\,H\,(d+a)\,(d+a-H)}{g\,(d+a)^2\,(d+a-H)\,-\,d^2\,c^2}. \label{relparasolsermod2}
\end{eqnarray}

\begin{figure}
  \centering
  \subfigure[Solitary wave]{\includegraphics[width=0.49\textwidth]{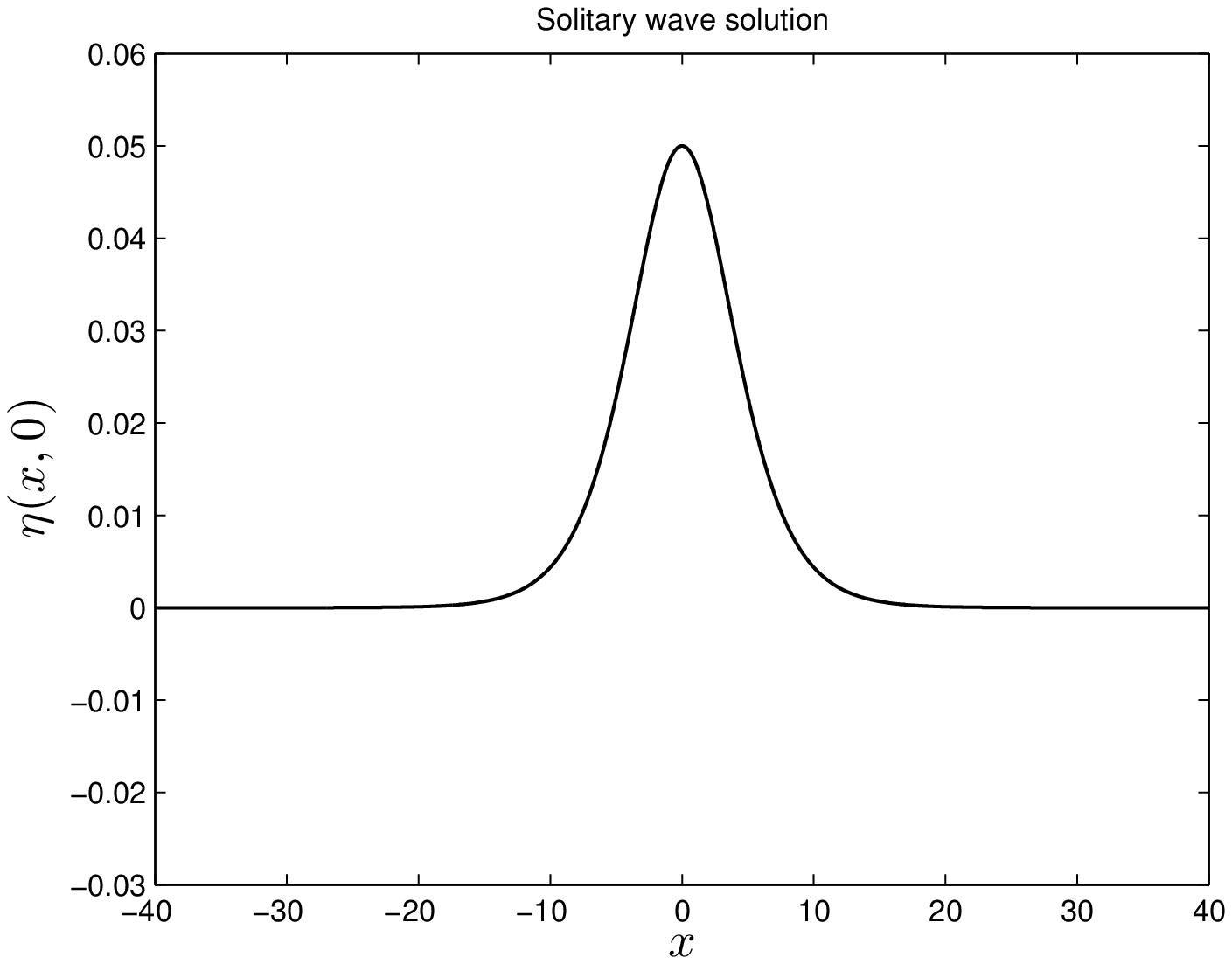}}
  \subfigure[Cnoidal wave]{\includegraphics[width=0.49\textwidth]{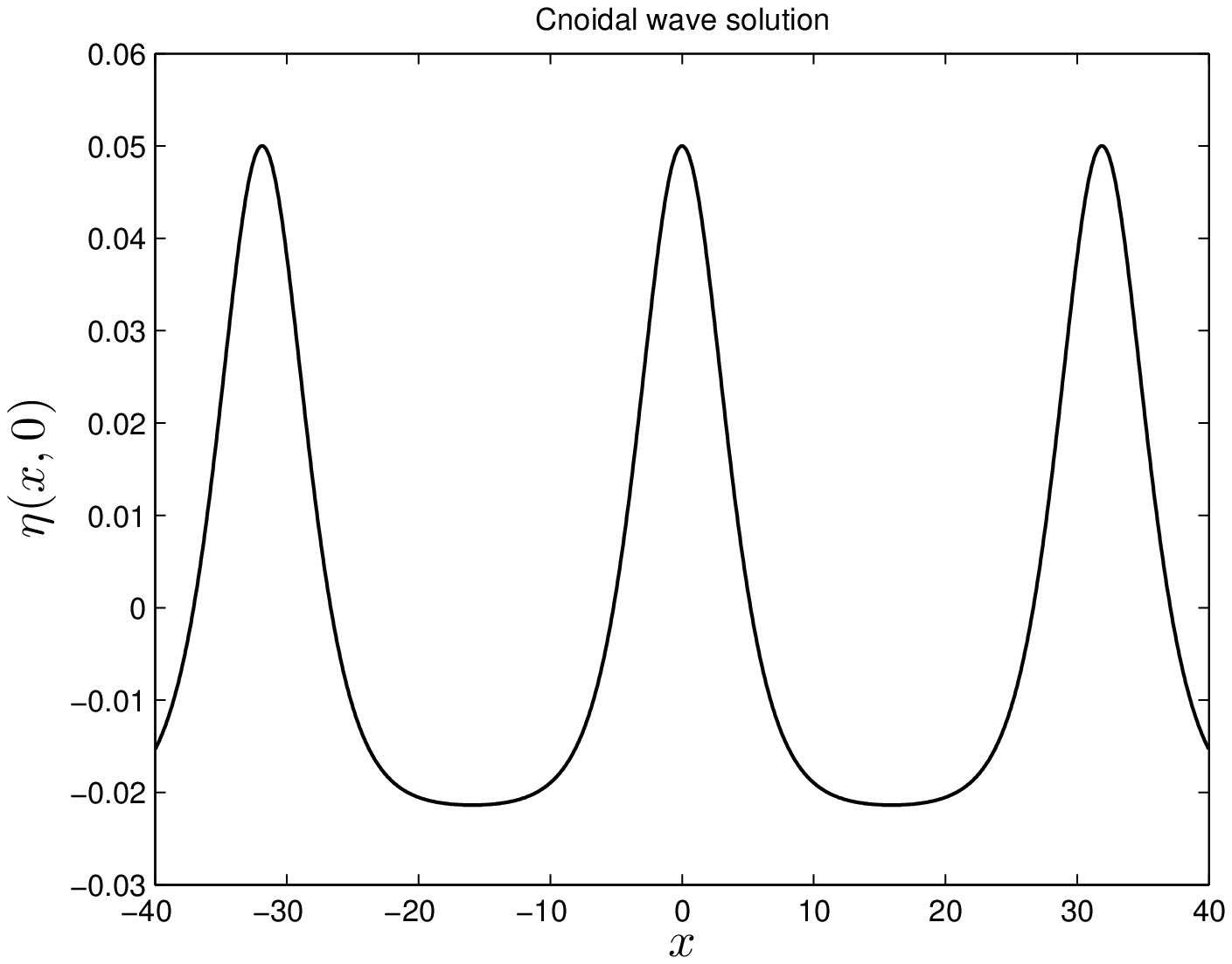}}
  \caption{\em Two exact solutions to the Serre equations. The solitary wave amplitude is equal to $a = 0.05$. For the cnoidal wave parameters $m$ and $a$ are equal to $0.99$ and $0.05$ respectively. Other cnoidal wave parameters are deduced from relations \eqref{relparasolsermod}, \eqref{relparasolsermod2}.}
  \label{fig:esols}
\end{figure}

However, in the present study, we are interested in the classical solitary wave solution which is recovered in the limiting case $m \rightarrow 1$
\begin{equation}\label{eq:swsol}
  \eta = a\ \sech^2\half \varkappa(x-ct), \quad
  \um = \frac{c\,\eta}{d+\eta}, \quad
  c^2 = g(d+a), \quad
  (\varkappa d)^2 = \frac{a}{\beta(d+a)}.
\end{equation}
For illustrative purposes, a solitary wave along with a cnoidal wave of the same amplitude $a = 0.05$ are depicted in Figure~\ref{fig:esols}.

\begin{figure}
  \centering
  \subfigure[$c = 1.1$]%
  {\includegraphics[width=0.49\textwidth]{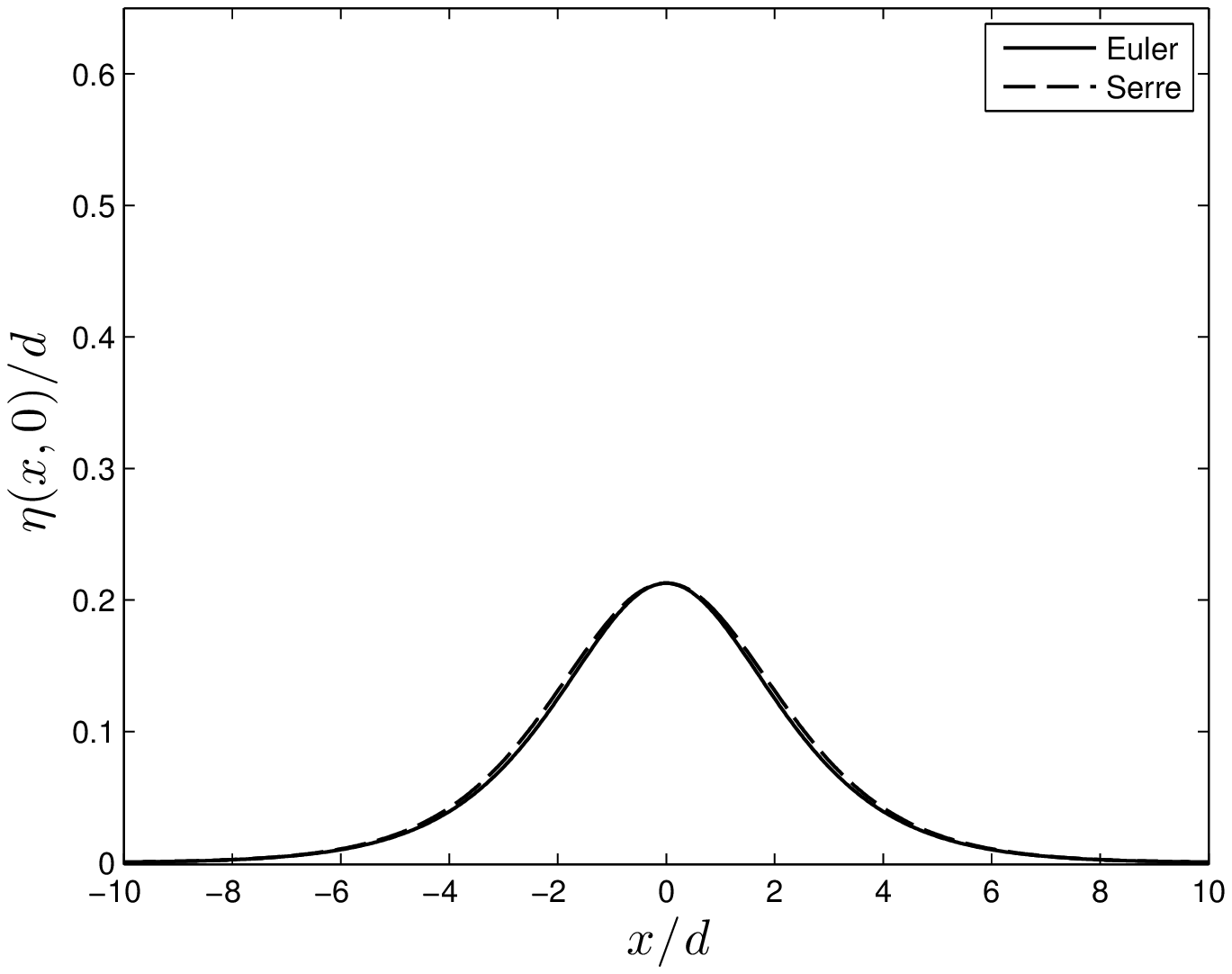}}
  \subfigure[$c = 1.15$]%
  {\includegraphics[width=0.49\textwidth]{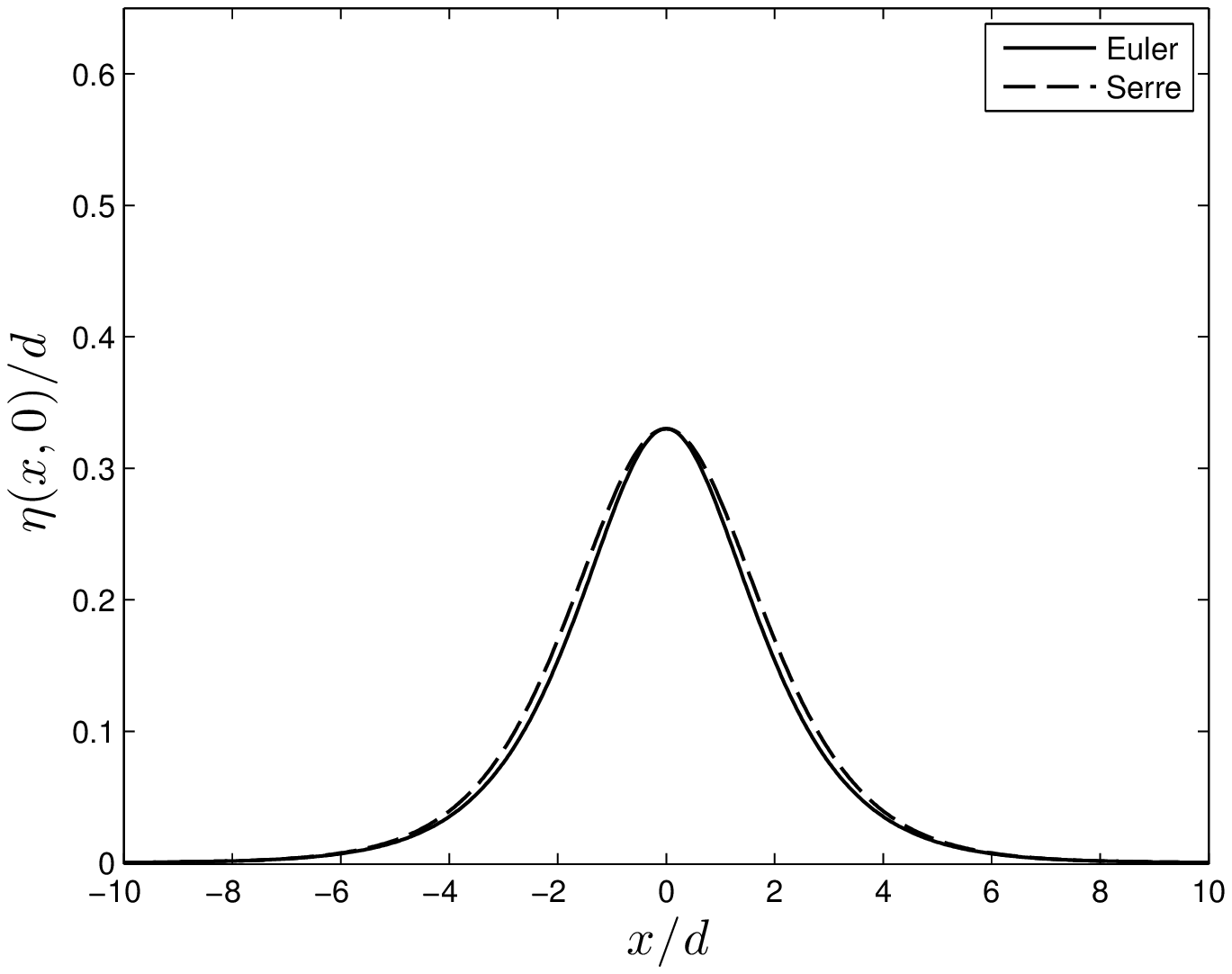}}
  \subfigure[$c = 1.2$]%
  {\includegraphics[width=0.49\textwidth]{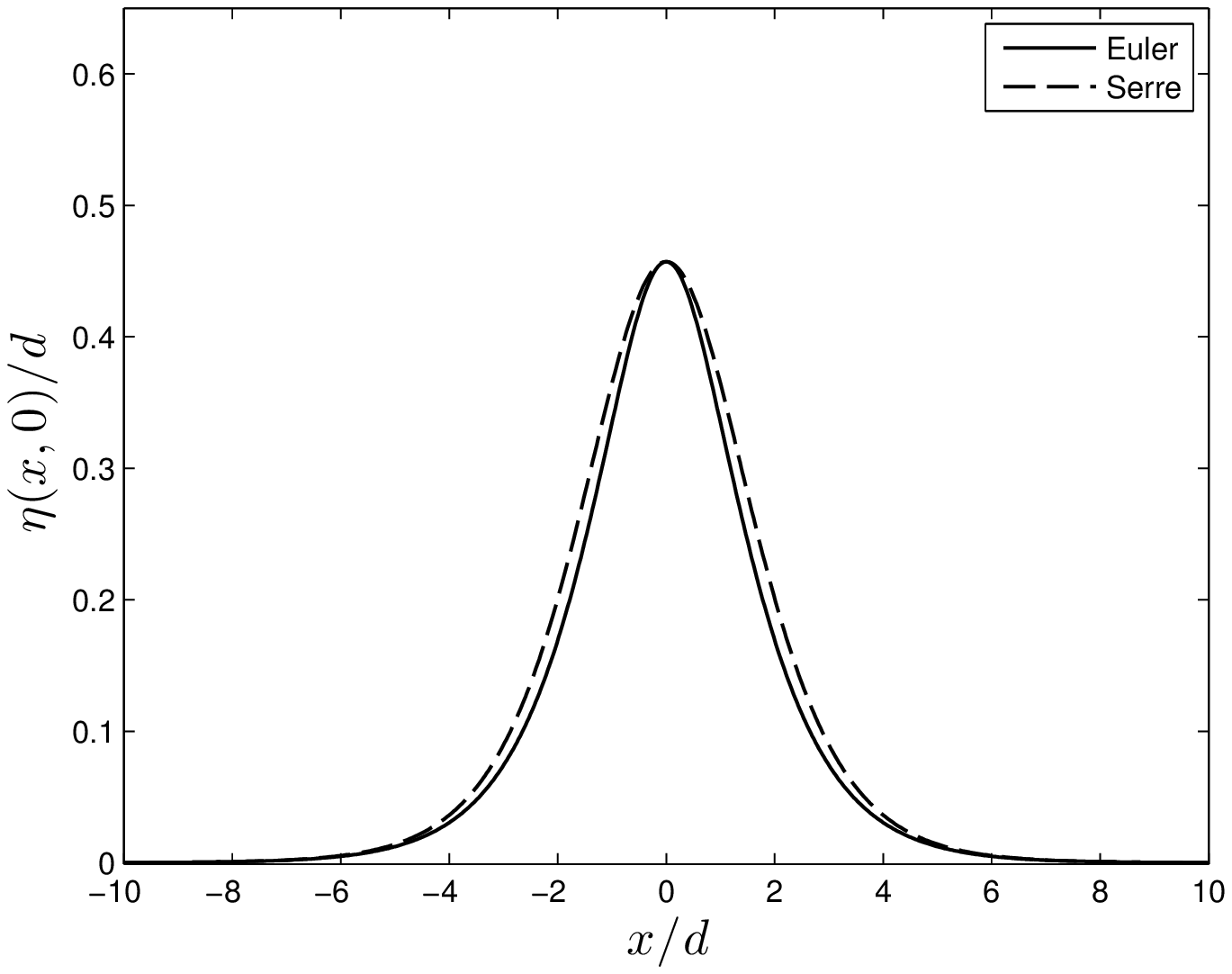}}
  \subfigure[$c = 1.25$]%
  {\includegraphics[width=0.49\textwidth]{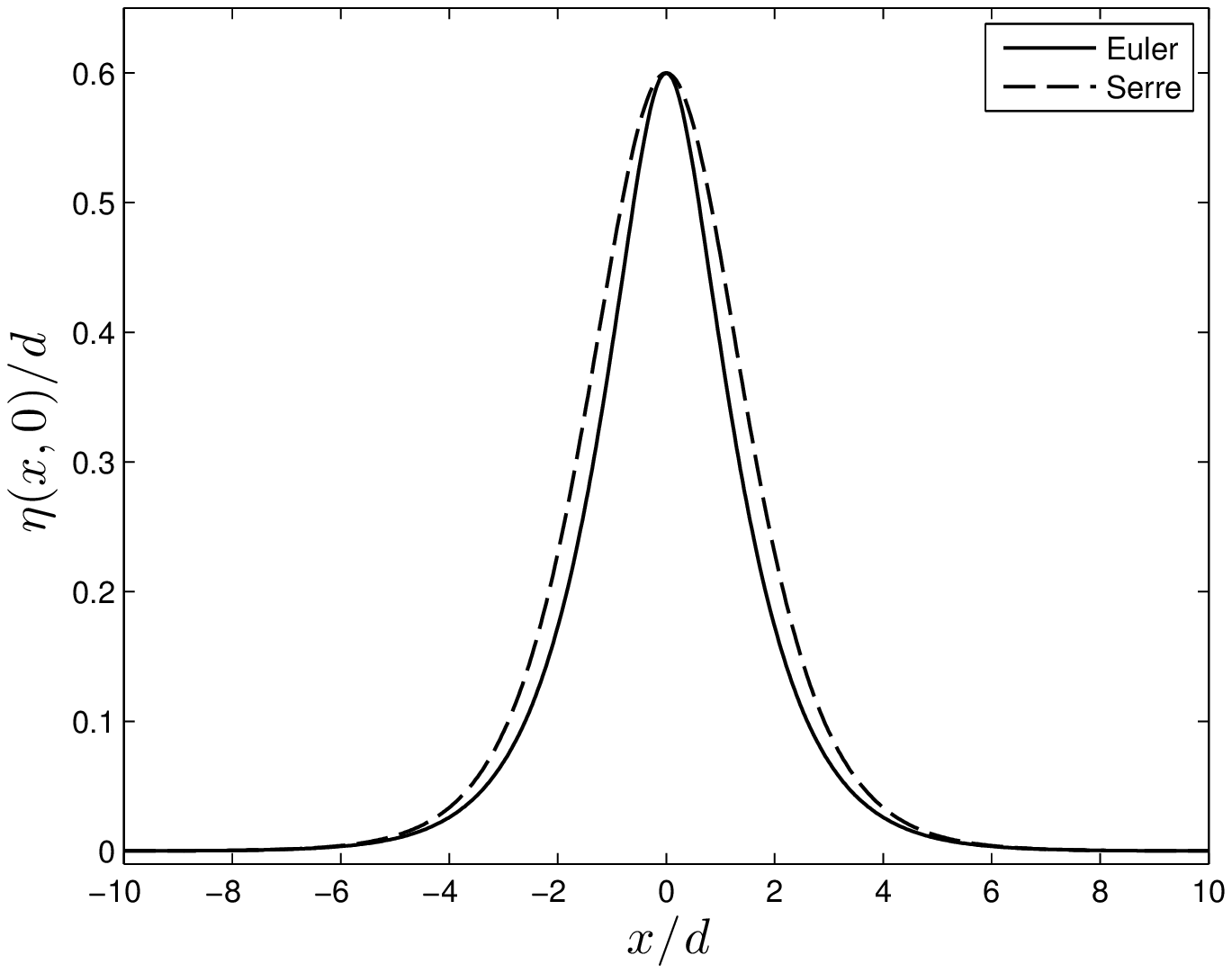}}
  \caption{\em Comparison of solitary wave solutions to the Serre and the full Euler equations.}
  \label{fig:sws}
\end{figure}

Using the exact solitary wave solution \eqref{eq:swsol} we can assess the accuracy of the Serre equations (with $\beta = \third$) by making comparisons with corresponding solutions to the original full Euler equations. The procedure we use to construct traveling wave solutions to Euler equations is described in \cite{Clamond2012b}. The \textsc{Matlab} script used to generate these profiles (up to machine precision) can be freely downloaded from the File Exchange server \cite{Clamond2012}. The results of comparison for several values of the speed parameter $c$ are presented on Figure~\ref{fig:sws}. We can see that solitary waves to the Serre equations approximate fairly well the full Euler solutions approximately up to the amplitude $a/d = \half$. We note that similar conclusions were obtained in a previous study by \textsc{Li} \emph{et al.} (2004) \cite{Li2004}.

\subsection{Invariants of the Serre equations}

Henceforth we consider only the two-dimensional case. As pointed out by Y.~\textsc{Li} (2002) \cite{Li2002}, the classical Serre equations possess a non-canonical Hamiltonian structure which can be easily generalized for the model \eqref{eq:sermod}, \eqref{eq:sermod2}
\begin{equation*}
  \begin{pmatrix}
    h_t \\
    \tilde{q}_t
  \end{pmatrix}\ =\ \mathbb{J}\scal\begin{pmatrix}
                     \displaystyle{\delta\,\H\,/\,\delta\/ \tilde{q}} \\
                     \displaystyle{\delta\,\H\,/\,\delta\/ h}
                   \end{pmatrix},
\end{equation*}
where the Hamiltonian functional $\H$ and the symplectic operator $\mathbb{J}$ are defined as
\begin{equation*}
  \H\ =\ \half\int_\R\left[\,h\,\um^2\, +\, \beta\, h^3\,\um_x^2\, +\, g\,\eta^2\,\right]\,\ud\/x, 
  \qquad
  \mathbb{J}\ =\ -\begin{bmatrix}
         h_x & 0 \\
         \,\tilde{q}_x + \tilde{q}\/\partial_x & h\partial_x
       \end{bmatrix}.
\end{equation*}
The variable $\tilde{q}$ is defined by
\begin{equation*}
  \tilde{q}\ \equiv\ h\,\um\ -\ \beta\,[\,h^3\,\um_x\,]_x.
\end{equation*}
The conservation of the quantity $\tilde{q}$ was established in equation \eqref{eq:vflux}.

According to \cite{Li2002}, one-parameter symmetry groups of Serre's equations include the space translation $(x+\eps,t,h,u)$, the time translation $(x,t+\eps,h,u)$, the Galilean boost $(x+\eps t, t,h,u+\eps)$ and the scaling $\ue^\eps(\ue^\eps x,t,\ue^\eps h,u)$. Using the first three symmetry groups and the symplectic operator $\mathbb{J}$, one may recover the following invariants:
\begin{equation}\label{eq:HQ}
  \Q\ =\ \int_{\R}\frac{\eta\,\tilde{q}}{d\,+\,\eta}\,\ud\/x, \qquad
  \H, \qquad
  \int_{\R}\left[\,t\,\tilde{q}\, -\, x\,\eta\,\right]\,\ud\/x.
\end{equation}
Obviously, the equation \eqref{eq:sermod} leads to an invariant closely related to the mass conservation property $\int_{\R} \eta\, \ud\/x$. The scaling does not yield any conserved quantity with respect to the symplectic operator $\mathbb{J}$. Below, we are going to use extensively the generalized energy $\H$ and the generalized momentum $\Q$ conservation to assess the accuracy of the numerical schemes in addition to the exact analytical solution \eqref{eq:swsol}.

\section{Finite volume scheme and numerical results}\label{sec:nums}

In the present study, we propose a finite volume discretization procedure \cite{Barth1994, Barth2004} for the Serre equations \eqref{eq:sermod}, \eqref{eq:sermod2} that we rewrite here as
\begin{align}
  h_t\ +\ [\,h\,u\,]_x\ =&\ 0, \label{eq:ser1} \\
  u_t\ +\ \left[\,\half\,u^2\, +\, g\,h\,\right]_x\ =&\ \beta\,h^{-1}\,\left[\,h^3\,(u_{xt}\, 
  +\, u\,u_{xx}\, -\,u_x^{\,2})\,\right]_x, \label{eq:ser2}
\end{align}
where the over-bars have been omitted for brevity. (In this section, over-bars denote quantities averaged over a cell, as explained below.)

We begin our presentation by a discretization of the hyperbolic part of the equations (which are simply the classical Saint-Venant equations) and then discuss the treatment of dispersive terms. The Serre equations can be formally put under the quasilinear form
\begin{equation}\label{eq:conslaw}
  \boldsymbol{V}_t\ +\ [\,\boldsymbol{F}(\boldsymbol{V})\,]_x\ =\ \boldsymbol{S}(\boldsymbol{V}),
\end{equation}
where $\boldsymbol{V}$, $\boldsymbol{F}(\boldsymbol{V})$ are the conservative variables and the advective flux function, respectively
\begin{equation*}
  \boldsymbol{V}\ \equiv\ \begin{pmatrix}
        h \\
        u
      \end{pmatrix}, \quad
  \boldsymbol{F}(\boldsymbol{V})\ \equiv\ \begin{pmatrix}
           h\,u \\
           \half\,u^2\, +\, g\,h
         \end{pmatrix}.
\end{equation*}
The source term $\boldsymbol{S}(\boldsymbol{V})$ denotes the right-hand side of \eqref{eq:ser1}, \eqref{eq:ser2} and thus, depends also on space and time derivatives of $\boldsymbol{V}$. The Jacobian of the advective flux $\boldsymbol{F} (\boldsymbol{V})$ can be easily computed
\begin{equation*}
  \A(\boldsymbol{V})\ =\ \pd{\boldsymbol{F}(\boldsymbol{V})}{\boldsymbol{V}}\ =\ 
  \begin{bmatrix}
    u & h \\
    g & u
  \end{bmatrix}.
\end{equation*}
The Jacobian $\A(\boldsymbol{V})$ has two distinctive eigenvalues
\begin{equation*}
  \lambda^\pm\ =\ u\ \pm\ c_s, \qquad c_s\ \equiv\ \sqrt{gh}.
\end{equation*}
The corresponding right and left eigenvectors are provided here
\begin{equation*}
  \mathbb{R}\ =\ \begin{bmatrix}
        h & -h \\
        c_s & c_s
  	  \end{bmatrix}, \qquad
  \mathbb{L}\ =\ \mathbb{R}^{-1}\ =\ \frac12
  		      \begin{bmatrix}
                 h^{-1} & c_s^{-1} \\
                 -h^{-1} & c_s^{-1}
               \end{bmatrix}.
\end{equation*}

We consider a partition of the real line $\R$ into cells (or finite volumes) $\C_i = [x_{i-\frac12}, x_{i+\frac12}]$ with cell centers $x_i = \half(x_{i-\frac12} + x_{i+\frac12})$ ($i\in\Z$). Let $\Delta x_i$ denotes the length of the cell $\C_i$. In the sequel we will consider only uniform partitions with $\Delta x_i$ = $\Delta x$, $\forall i\in\Z$. We would like to approximate the solution $\boldsymbol{V}(x,t)$ by discrete values. In order to do so, we introduce the cell average of $\boldsymbol{V}$ on the cell $\C_i$ (denoted with an overbar), i.e.,
\begin{equation*}
\bar{\boldsymbol{V}}_i(t)\ \equiv\ \left(\,\hm_i(t)\,,\,\um_i(t)\,\right)\ =\ 
\frac{1}{\Delta x} \int_{\C_i} \boldsymbol{V}(x,t)\,\ud\/x.
\end{equation*}
A simple integration of \eqref{eq:conslaw} over the cell $\C_i$ leads the following exact relation:
\begin{equation*}
  \od{\bar{\boldsymbol{V}}}{t}\ +\ \frac{1}{\Delta x}\left[\,\boldsymbol{F}(\boldsymbol{V}(x_{i+\frac12},t))
  \, -\, \boldsymbol{F}(\boldsymbol{V}(x_{i-\frac12},t))\,\right]\ =\  
  \frac{1}{\Delta x}\int_{\C_i}\boldsymbol{S}(\boldsymbol{V})\,\ud\/x\ \equiv\ \bar{\boldsymbol{S}}_i.
\end{equation*}
Since the discrete solution is discontinuous at cell interfaces $x_{i+\frac12}$ ($i\in\Z$), we replace the flux at the cell faces by the so-called numerical flux function
\begin{equation*}
  \boldsymbol{F}(\boldsymbol{V}(x_{i\pm\frac12},t))\ \approx\ \F_{i\pm\frac12}(
  \bar{\boldsymbol{V}}_{i\pm\frac12}^{L},\bar{\boldsymbol{V}}_{i\pm\frac12}^{R}),
\end{equation*}
where $\bar{\boldsymbol{V}}_{i\pm\frac12}^{L,R}$ denotes the reconstructions of the conservative variables $\bar{\boldsymbol{V}}$ from left and right sides of each cell interface (the reconstruction procedure employed in the present study will be described below). Consequently, the semi-discrete scheme takes the form
\begin{equation}\label{eq:si1}
\od{\bar{\boldsymbol{V}}_i}{t}\ +\ \frac{1}{\Delta x}\left[\,\F_{i+\frac12}\, -\, 
\F_{i-\frac12}\,\right]\ =\ \bar{\boldsymbol{S}}_i.
\end{equation}

In order to discretize the advective flux $\boldsymbol{F}(\boldsymbol{V})$, we use the FVCF scheme \cite{Ghidaglia1996, Ghidaglia2001}:
\begin{equation*}
\F(\boldsymbol{V},\boldsymbol{W})\ =\ \frac{\boldsymbol{F}(\boldsymbol{V})\,+\,
\boldsymbol{F}(\boldsymbol{W})}{2}\ -\ \mathbb{U}(\boldsymbol{V},\boldsymbol{W})\scal
\frac{\boldsymbol{F}(\boldsymbol{W})\,-\,\boldsymbol{F}(\boldsymbol{V})}{2}.
\end{equation*}
The first part of the numerical flux is centered, the second part is the upwinding introduced through the Jacobian sign-matrix $\mathbb{U}(\boldsymbol{V},\boldsymbol{W})$ defined as
\begin{equation*}
  \mathbb{U}(\boldsymbol{V},\boldsymbol{W})\ =\ \sign\left[\A\!\left(\half(\boldsymbol{V}
  +\boldsymbol{W})\right)\right], \qquad 
  \sign(\A)\ =\ \mathbb{R}\scal\diag(s^+, s^-)\scal\mathbb{L},
\end{equation*}
where $s^\pm \equiv \sign(\lambda^\pm)$. After some simple algebraic computations, one can find
\begin{equation*}
  \mathbb{U}\ =\ \frac{1}{2}\begin{bmatrix}
   		s^+ + s^- & (h/c_s)\,(s^+ - s^-) \\
   		(g/c_s)\,(s^+ - s^-) & s^+ + s^-
      \end{bmatrix},
\end{equation*}
the sign-matrix $\mathbb{U}$ being evaluated at the average state of left and right values.

\subsection{High order reconstruction}

In order to obtain a higher-order scheme in space, we need to replace the piecewise constant data by a piecewise polynomial representation. This goal is achieved by various so-called reconstruction procedures such as MUSCL TVD \cite{Kolgan1975, Leer1979, Leer2006}, UNO \cite{HaOs}, ENO \cite{Harten1989}, WENO \cite{Xing2005} and many others. In our previous study on Boussinesq-type equations \cite{Dutykh2011e}, the UNO2 scheme showed a good performance with small dissipation in realistic propagation and run-up simulations. Consequently, we retain this scheme for the discretization of the advective flux in Serre equations.

\begin{remark}
In TVD schemes, the numerical operator is required (by definition) not to increase the total variation of the numerical solution at each time step. It follows that the value of an isolated maximum may only decrease in time which is not a good property for the simulation of coherent structures such as solitary waves. The non-oscillatory UNO2 scheme, employed in our study, is only required to diminish the \emph{number} of local extrema in the numerical solution. Unlike TVD schemes, UNO schemes are not constrained to damp the values of each local extremum at every time step.
\end{remark}

The main idea of the UNO2 scheme is to construct a non-oscillatory piecewise-parabolic interpolant $\boldsymbol{Q}(x)$ to a piecewise smooth function $\boldsymbol{V}(x)$ (see \cite{HaOs} for more details). On each segment containing the face $x_{i+\frac12} \in [x_i, x_{i+1}]$, the function $\boldsymbol{Q}(x) = \boldsymbol{q}_{i+\frac12}(x)$ is locally a quadratic polynomial and wherever $v(x)$ is smooth we have
\begin{equation*}
  \boldsymbol{Q}(x)\ -\ \boldsymbol{V}(x)\ =\ \boldsymbol{0}\ +\ \O(\Delta x^3), \qquad
  \od{\boldsymbol{Q}}{x}(x\pm 0)\ -\ \od{\boldsymbol{V}}{x}\ =\ \boldsymbol{0}\ +\ \O(\Delta x^2).
\end{equation*}
Also, $\boldsymbol{Q}(x)$ should be non-oscillatory in the sense that the number of its local extrema does not exceed that of $\boldsymbol{V}(x)$. Since $\boldsymbol{q}_{i+\frac12}(x_i) = \bar{\boldsymbol{V}}_i$ and $\boldsymbol{q}_{i+\frac12}(x_{i+1}) = \bar{\boldsymbol{V}}_{i+1}$, it can be written in the form
\begin{equation*}
  \boldsymbol{q}_{i+\frac12}(x)\ =\ \bar{\boldsymbol{V}}_i\ +\ 
  \mathfrak{d}_{i+\frac12}\{\boldsymbol{V}\}\times\frac{x - x_i}{\Delta x}\ +\   \half\,\mathfrak{D}_{i+\frac12}\{\boldsymbol{V}\}\times\frac{(x-x_i)(x-x_{i+1})}{\Delta x^2},
\end{equation*}
where $\mathfrak{d}_{i+\frac12} \{\boldsymbol{V}\} \equiv\bar{\boldsymbol{V}}_{i+1} - \bar{\boldsymbol{V}}_{i}$ and $\mathfrak{D}_{i+\frac12} \boldsymbol{V}$ is closely related to the second derivative of the interpolant since $\mathfrak{D}_{i+\frac12} \{\boldsymbol{V}\} = \Delta x^2\, \boldsymbol{q}''_{i+\frac12}(x)$. The polynomial $\boldsymbol{q}_{i+\frac12} (x)$ is chosen to be the least oscillatory between two candidates interpolating $\boldsymbol{V}(x)$ at $(x_{i-1}, x_i, x_{i+1})$ and $(x_i, x_{i+1}, x_{i+2})$. This requirement leads to the following choice of $\mathfrak{D}_{i+\frac12} \{\boldsymbol{V}\} \equiv \minmod\left(\mathfrak{D}_i \{\boldsymbol{V}\}, \mathfrak{D}_{i+1} \{\boldsymbol{V}\}\right)$ with
\begin{equation*}
  \mathfrak{D}_i\{\boldsymbol{V}\}\ =\ \bar{\boldsymbol{V}}_{i+1}\ -\ 2\,
  \bar{\boldsymbol{V}}_i\ +\ \bar{\boldsymbol{V}}_{i-1}, \qquad
  \mathfrak{D}_{i+1}\{\boldsymbol{V}\}\ =\ \bar{\boldsymbol{V}}_{i+2}\ -\ 2\,\bar{\boldsymbol{V}}_{i+1} \ +\ \bar{\boldsymbol{V}}_i,
\end{equation*}
and where $\minmod(x,y)$ is the usual minmod function defined as
\begin{equation*}
  \minmod (x,y)\ \equiv\ \half\,[\,\sign(x)\,+\,\sign(y)\,]\times\min(|x|,|y|).
\end{equation*}

To achieve the second order $\O(\Delta x^2)$ accuracy, it is sufficient to consider piecewise linear reconstructions in each cell. Let $L(x)$ denote this approximately reconstructed function which can be written in this form
\begin{equation*}
  L(x)\ =\ \bar{\boldsymbol{V}}_i\ +\ \boldsymbol{S}_i\times\frac{x-x_i}{\Delta x}, \qquad x \in [x_{i-\frac12}, x_{i+\frac12}].
\end{equation*}
In order to $L(x)$ be a non-oscillatory approximation, we use the parabolic interpolation $\boldsymbol{Q}(x)$ constructed below to estimate the slopes $\boldsymbol{S}_i$ within each cell
\begin{equation*}
\boldsymbol{S}_i\ =\ \Delta x\times\minmod\left(\od{\boldsymbol{Q}}{x}(x_i-0),
\od{\boldsymbol{Q}}{x}(x_i+0)\right).
\end{equation*}
In other words, the solution is reconstructed on the cells while the solution gradient is estimated on the dual mesh as it is often performed in more modern schemes \cite{Barth1994, Barth2004}. A brief summary of the UNO2 reconstruction can be also found in \cite{Dutykh2011e, Dutykh2010e}.

\subsection{Treatment of the dispersive terms}

In this section, we explain how we treat the dispersive terms of Serre equations \eqref{eq:ser1}, \eqref{eq:ser2}. We begin the exposition by discussing the space discretization and then, we propose a way to remove the intrinsic stiffness of the dispersion by partial implicitation.

For the sake of simplicity, we split the dispersive terms into three parts:
\begin{align*}
  \M(\boldsymbol{V})\ \equiv\ \beta\,h^{-1}\,\left[\,h^3\,u_{xt}\,\right]_x, \quad
  \D_1(\boldsymbol{V})\ \equiv\ \beta\, h^{-1}\,\left[\,h^3\,u\,u_{xx}\,\right]_x, \quad
  \D_2(\boldsymbol{V})\ \equiv\ \beta\, h^{-1}\,\left[\,h^3\,u_x^{\,2}\,\right]_x.
\end{align*}
We propose the following approximations in space (which are all of the second order $\O(\Delta x^2)$ to be consistent with UNO2 advective flux discretization presented above)
\begin{align*}
  \M_i(\bar{\boldsymbol{V}})\ &=\ \beta\,\hm_i^{\,-1}\,\frac{\hm_{i+1}^{\,3}\,(\um_{xt})_{i+1}\, 
  -\, \hm_{i-1}^{\,3}\,(\um_{xt})_{i-1}}{2\,\Delta x} \\ 
  &=\ \frac{\beta\, \hm_i^{\,-1}}{2\,\Delta x}\,\left[\,
  \hm_{i+1}^{\,3}\,\frac{(\um_t)_{i+2}\, -\, (\um_t)_i}{2\,\Delta x}\, -\, 
  \hm_{i-1}^{\,3}\,\frac{(\um_t)_i\, -\, (\um_t)_{i-2}}{2\,\Delta x}\,\right] \\ 
  &=\ \frac{\beta\,\hm_i^{\,-1}}{4\,\Delta x^2}\,\left[\,
  \hm_{i+1}^{\,3}\,(\um_t)_{i+2}\, -\, (\hm_{i+1}^{\,3}\, +\, \hm_{i-1}^{\,3})\,(\um_t)_i\, +\,
  \hm_{i-1}^{\,3}\,(\um_t)_{i-2}\,\right].
\end{align*}
The last relation can be rewritten in a short-hand form if we introduce the matrix $\M(\bar{\boldsymbol{V}})$ such that the $i$-th component of the product $\M(\bar{\boldsymbol{V}}) \scal \bar{\boldsymbol{V}}_t$ gives exactly the expression $\M_i(\bar{\boldsymbol{V}})$.

In a similar way, we discretize the other dispersive terms without giving here the intermediate steps
\begin{align*}
  \D_{1i}(\bar{\boldsymbol{V}})\ &=\ \frac{\beta\, \hm_i^{-1}}{2\,\Delta x^3}\,\left[\,
  \hm_{i+1}^{\,3}\,\um_{i+1}\,(\um_{i+2} - 2\um_{i+1} + \um_i)\, -\,
  \hm_{i-1}^{\,3}\,\um_{i-1}\,(\um_i - 2\um_{i-1} + \um_{i-2})\,\right],\\
  \D_{2i}(\bar{\boldsymbol{V}})\ &=\ \frac{\beta \hm_i^{\,-1}}{8\Delta x^3}\,\left[\,
  \hm_{i+1}^{\,3}\,(\um_{i+2} - \um_i)^2\, -\, \hm_{i-1}^{\,3}\,(\um_i - \um_{i-2})^2\,\right].
\end{align*}
In a more general nonperiodic case asymmetric finite differences should be used near the boundaries. If we denote by $\I$ the identity matrix, we can rewrite the semi-discrete scheme \eqref{eq:si1} by expanding the right-hand side $\boldsymbol{S}_i$
\begin{align}\label{eq:sm1}
  \od{\hm}{t}\ +\ \frac{1}{\Delta x}\,\left[\,\mathcal{F}_{+}^{(1)}(\bar{\boldsymbol{V}}) 
  \,-\, \mathcal{F}_{-}^{(1)}(\bar{\boldsymbol{V}})\,\right]\ &=\ 0, \\
  (\I - \M)\cdot\od{\um}{t}\ +\ \frac{1}{\Delta x}\,\left[\,
  \mathcal{F}_{+}^{(2)}(\bar{\boldsymbol{V}})
  \,-\, \mathcal{F}_{-}^{(2)}(\bar{\boldsymbol{V}})\,\right]\ &=\ \D(\bar{\boldsymbol{V}})\cdot\um, \label{eq:sm2}
\end{align}
where $\mathcal{F}_{\pm}^{(1,2)}(\bar{\boldsymbol{V}})$ are the two components of the advective numerical flux vector $\F$ at the right ($+$) and left ($-$) faces correspondingly and $\D(\bar{\boldsymbol{V}}) \equiv \D_1(\bar{\boldsymbol{V}}) - \D_2(\bar{\boldsymbol{V}})$.

Finally, in order to obtain the semidiscrete scheme, one has to solve a linear system to find explicitly the time derivative $\text{d}\/\um/\text{d}t$. A mathematical study of the resulting matrix $\I-\M$ is not straightforward to perform. However, in our numerical tests we have never experienced any difficulties to invert it.

\subsection{Temporal scheme}

We rewrite the inverted semi-discrete scheme \eqref{eq:sm1}--\eqref{eq:sm2} as a system of ODEs:
\begin{equation*}
\partial_t\,w\ =\ \L(w, t), \qquad w(0)\ =\ w_0.
\end{equation*}
In order to solve numerically the last system of equations, we apply the Bogacki--Shampine method \cite{Bogacki1989}. It is a third-order Runge--Kutta scheme with four stages. It has an embedded second-order method which is used to estimate the local error and, thus, to adapt the time step size. Moreover, the Bogacki--Shampine method enjoys the First Same As Last (FSAL) property so that it needs three function evaluations per step. This method is also implemented in the \texttt{ode23} function in {\sc Matlab} \cite{Shampine1997}. A step of the Bogacki--Shampine method is given by
\begin{align*}
k_1\ &=\ \L(w^{(n)},t_n), \\
k_2\ &=\ \L(w^{(n)}+\half \Delta t_n k_1, t_n + \half\Delta t), \\
k_3\ &=\ \L(w^{(n)})+\frth \Delta t_n k_2, t_n + \frth\Delta t), \\
w^{(n+1)}\ &=\ w^{(n)}\ +\ \Delta t_n\times\left(\textstyle{2\over9}k_1 + 
\textstyle{1\over3}k_2 + \textstyle{4\over9}k_3\right), \\
k_4\ &=\ \L(w^{(n+1)}, t_n + \Delta t_n), \\
w_2^{(n+1)}\ &=\ w^{(n)}\ +\ \Delta t_n\times\left(\textstyle{4\over{24}}k_1 + 
\textstyle{1\over4}k_2 + \textstyle{1\over3} k_3 + \textstyle{1\over8}k_4\right).
\end{align*}
Here $w^{(n)}\approx w(t_n)$, $\Delta t$ is the time step and $w_2^{(n+1)}$ is a second order approximation to the solution $w(t_{n+1})$, so the difference between $w^{(n+1)}$ and $w_2^{(n+1)}$ gives an estimation of the local error. The FSAL property consists in the fact that $k_4$ is equal to $k_1$ in the next time step, thus saving one function evaluation.

If the new time step $\Delta t_{n+1}$ is given by $\Delta t_{n+1} = \rho_n\Delta t_n$, then according to H211b digital filter approach \cite{Soderlind2003, Soderlind2006}, the proportionality factor $\rho_n$ is given by:
\begin{equation}\label{eq:tadapt}
\rho_n\ =\,\left(\frac{\delta}{\eps_n}\right)^{\!\beta_1}\left(\frac{\delta}
{\eps_{n-1}}\right)^{\!\beta_2}\,\rho_{n-1}^{\,-\alpha},
\end{equation}
where $\eps_n$ is a local error estimation at time step $t_n$, $\delta$ is the desired tolerance and the constants $\beta_1$, $\beta_2$ and $\alpha$ are defined as
\begin{equation*}
\alpha\ =\ \frac{1}{4}, \qquad \beta_1\ =\ \beta_2\ =\ \frac{1}{4\,p}.
\end{equation*}
The parameter $p$ is the order of the scheme ($p=3$ in our case).

\begin{remark}
The adaptive strategy \eqref{eq:tadapt} can be further improved if we smooth the factor $\rho_n$ before computing the next time step $\Delta t_{n+1}$
\begin{equation*}
\Delta t_{n+1}\ =\ \hat{\rho}_n\,\Delta t_n, \qquad
\hat{\rho}_n\ =\ \omega(\rho_n).
\end{equation*}
The function $\omega(\rho)$ is called \emph{the time step limiter} and should be smooth, monotonically increasing and should satisfy the following conditions
\begin{equation*}
\omega(0)\ <\ 1, \qquad \omega(+\infty)\ >\ 1, \qquad \omega(1)\ =\ 1, \qquad \omega'(1)\ =\ 1.
\end{equation*}
One possible choice is suggested in \cite{Soderlind2006}:
\begin{equation*}
\omega(\rho)\ =\ 1\ +\ \kappa\/\arctan\!\left(\frac{\rho-1}{\kappa}\right).
\end{equation*}
In our computations the parameter $\kappa$ is set to 1.
\end{remark}

\section{Pseudo-spectral Fourier-type method for the Serre equations}\label{sec:spectral}

In this Section we describe a pseudo-spectral solver to integrate numerically the Serre equations in periodic domains. In spectral methods, it is more convenient to take as variables the free surface elevation $\eta(x,t)$ and the conserved quantity $q(x,t)$
\begin{align}
  \eta_t\ +\ [\,(d+\eta)\,\um\,]_x\ &=\ 0, \label{eq:s1} \\
  q_t\ +\ \left[\,q\,u\, -\,\half\,\um^2\, +\, g\,\eta\, -\, \half\,(d+\eta)^2\,\um_x^2\,\right]_x\ &=\ 0, 
  \label{eq:s2} \\
  q\ -\ \um\ +\ \third(d+\eta)^2\um_{xx}\ +\ (d+\eta)\eta_x\um_x\ &=\ 0. \label{eq:s3}
\end{align}
The first two equations \eqref{eq:s1}, \eqref{eq:s2} are of evolution type, while the third one \eqref{eq:s3} relates the conserved variable $q$ to the primitive variables: the free surface elevation $\eta$ and the velocity $\um$. In order to solve relation \eqref{eq:s3} with respect to the velocity $\um$, we extract the linear part as
\begin{equation*}
  \um\ -\ \third\,d^2\,\um_{xx}\ -\ q\ =\ \underbrace{\,\third\,(2d\eta + \eta^2)\,\um_{xx}\ +\ 
  (d+\eta)\,\eta_x\,\um_x}_{N(\eta,\um)\,}.
\end{equation*}
Then, we apply to the last relation the following fixed point type iteration in Fourier space
\begin{equation}\label{eq:fixed}
  \hat{\um}_{j+1}\ =\ \frac{\hat{q}}{1+\third(kd)^2}\ +\ \frac{\F\left\{N(\eta,\um_j)\right\}}
  {1+\third(kd)^2} \qquad j=0,1,2,\cdots,
\end{equation}
where $\hat{\psi} \equiv \F\{\psi\}$ denotes the Fourier transform of a quantity $\psi$. The last iteration is repeated until the desired convergence. For example, for moderate amplitude solitary waves ($\approx 0.2$), the accuracy $10^{-16}$ is attained in approximatively 20 iterations if the velocity $\um_0$ is initialized from the previous time step. We note that the usual $3/2$ rule is applied to the nonlinear terms for anti-aliasing \cite{Trefethen2000, Clamond2001, Fructus2005}.

\begin{remark}
One can improve the fixed point iteration \eqref{eq:fixed} by employing the so-called relaxation approach \cite{Isaacson1966}. The relaxed scheme takes the following form
\begin{equation*}
  \hat{\um}_{j+1}\ =\,\left(\frac{\hat{q}}{1+\third(kd)^2}\ +\ \frac{\F\left\{N(\eta,\um_j)\right\}}
  {1+\third(kd)^2}\right)\theta\ +\ (1-\theta)\hat{\um}_{j} \qquad j=0,1,2,\cdots,
\end{equation*}
where $\theta\in [0,1]$ is a free parameter. We obtained the best convergence rate for $\theta = \half$.
\end{remark}

In order to improve the numerical stability of the time stepping method, we will integrate exactly the linear terms in evolution equations
\begin{align*}
  \eta_t\ +\ d\,\um_x\ &=\ -\/[\,\eta\,\um\,]_x, \\
  q_t\ +\ g\,\eta_x\ &=\ \left[\,\half\,\um^2\, +\, \half\,(d+\eta)^2\,\um_x^2\, -\, q\,u\,\right]_x.
\end{align*}
Taking the Fourier transform and using the relation \eqref{eq:s3} between $\um$ and $q$, we obtain the following system of ODEs:
\begin{align*}
  \hat{\eta}_t\ +\ \frac{\ui\/k\/d}{1+\third(kd)^2}\,\hat{q}\ &=\ -\/\ui\/k\,\F\{\eta\um\}\ 
  -\ \frac{\ui\/k\/d\,\F\left\{N(\eta,\um_j)\right\}}{1+\third(kd)^2}, \\
  \hat{q}_t\ +\ \ui\/k\/g\,\hat{\eta}\ &=\ \ui\/k\,\F\left\{\,
  \half\um^2\, +\, \half(d+\eta)^2\um_x^2\, -\, qu\,\right\}.
\end{align*}
The next step consists in introducing the vector of dimensionless variables in Fourier space $\hat{\boldsymbol V} \equiv (\ui\/k\/\hat{\eta}, \ui\/\omega\/\hat{q}/g)$, where $\omega^2=gk^2d/[1+\third(kd)^2]$ is the dispersion relation of the linearized Serre equations. With unscaled variables in vectorial form, the last system becomes
\begin{equation*}
  \hat{\boldsymbol V}_t\ +\ \L\scal\hat{\boldsymbol V}\ =\ \N(\hat{\boldsymbol V}), \qquad
  \L\ \equiv\ \begin{bmatrix}
          0 & \ui\/\omega \\
          \ui\/\omega & 0
        \end{bmatrix}.
\end{equation*}
On the right-hand side, we put all the nonlinear terms
\begin{equation*}
  \N(\hat{\boldsymbol V})\ =\ \begin{pmatrix}
    k^2\,\F\{\eta\um\}\, +\, d\/k^2\,\F\left\{N(\eta,\um_j)\right\}/(1+\third(kd)^2) \\
    -\/(k\omega/g)\,\F\left\{\half\um^2\, +\, \half(d+\eta)^2\um_x^2\, -\, qu\right\}
  \end{pmatrix}.
\end{equation*}
In order to integrate the linear terms, we make a last change of variables \cite{Milewski1999, Fructus2005}:
\begin{equation*}
  \hat{\boldsymbol W}_t\ =\ \ue^{(t-t_0)\L}\scal\N\left\{\ue^{-(t-t_0)\L}\scal\hat{\boldsymbol W}\right\},
   \qquad 
  \hat{\boldsymbol W}(t)\ \equiv\ \ue^{(t-t_0)\L}\scal\hat{\boldsymbol V}(t), \qquad 
  \hat{\boldsymbol W}(t_0)\ =\ \hat{\boldsymbol V}(t_0).
\end{equation*}
Finally, the last system of ODEs is discretized in time by Verner's embedded adaptive 9(8) Runge--Kutta scheme \cite{Verner1978}. The time step is chosen adaptively using the so-called \textsc{H211b} digital filter \cite{Soderlind2003, Soderlind2006} to meet some prescribed error tolerance (generally of the same order of the fixed point iteration \eqref{eq:fixed} precision). Since the numerical scheme is implicit in the velocity variable $\um$, the resulting time step $\Delta t$ is generally of the order of the spatial discretization $\O(\Delta x)$.

\section{Numerical results}\label{sec:numres}

In this section we present some numerical results using the finite volume scheme described hereinabove. First, we validate the discretization and check the convergence of the scheme using an analytical solution. Then we demonstrate the ability of the scheme to simulate the practically important solitary wave interaction problem. Throughout this section we consider the initial value problem with periodic boundary conditions unless a special remark is made.

\subsection{Convergence test and invariants preservation}

Consider the Serre equations \eqref{eq:ser1}, \eqref{eq:ser2} posed in the periodic domain $[-40, 40]$. We solve numerically the initial-periodic boundary value problem with an exact solitary wave solution \eqref{eq:swsol} posed as an initial condition. Then, this specific initial disturbance will be translated in space with known celerity under the system dynamics. This particular class of solutions plays an important role in water wave theory \cite{John, DMS1} and it will allow us to assess the accuracy of the proposed scheme. The values of the various physical parameters used in the simulation are given in Table~\ref{tab:params}.

\begin{table}
  \centering
	\begin{tabular}{l|c}
	  \hline\hline
	  Undisturbed water depth: $d$ & 1 \\
	  Gravity acceleration: $g$ & 1 \\
	  Solitary wave amplitude: $a$ & 0.05 \\
	  Final simulation time: $T$ & 2 \\
	  Free parameter: $\beta$ & 1/3 \\
	  \hline\hline
	\end{tabular}
  \caption{\em Values of various parameters used in convergence tests.}
  \label{tab:params}
\end{table}

\begin{figure}
  \centering
  \includegraphics[width=0.7\textwidth]{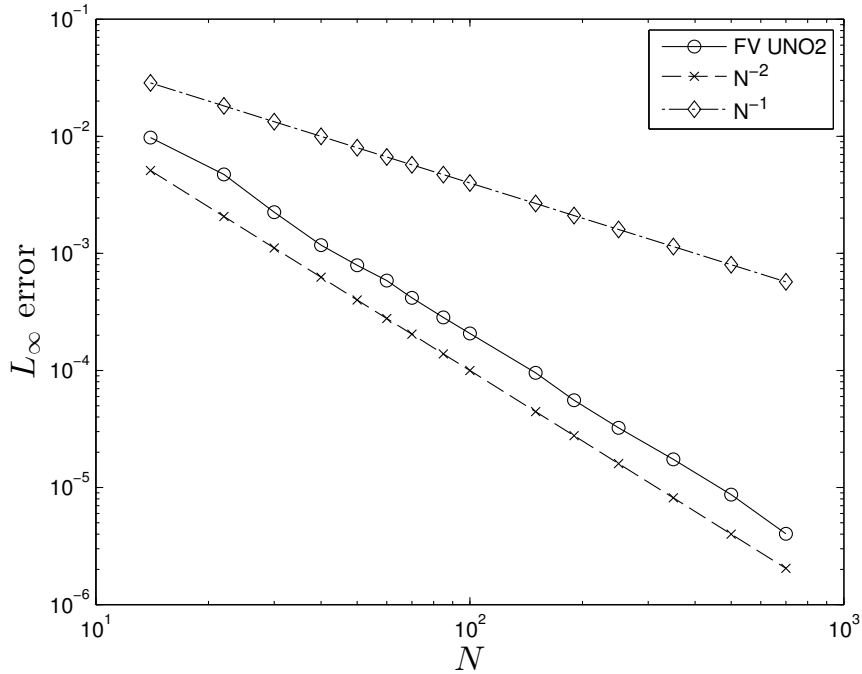}
  \caption{\em Convergence of the numerical solution in the $L_\infty$ norm computed using the finite volume method.}
  \label{fig:conv}
\end{figure}

\begin{figure}
  \centering
  \subfigure[Hamiltonian $\H$]%
  {\includegraphics[width=0.49\textwidth]{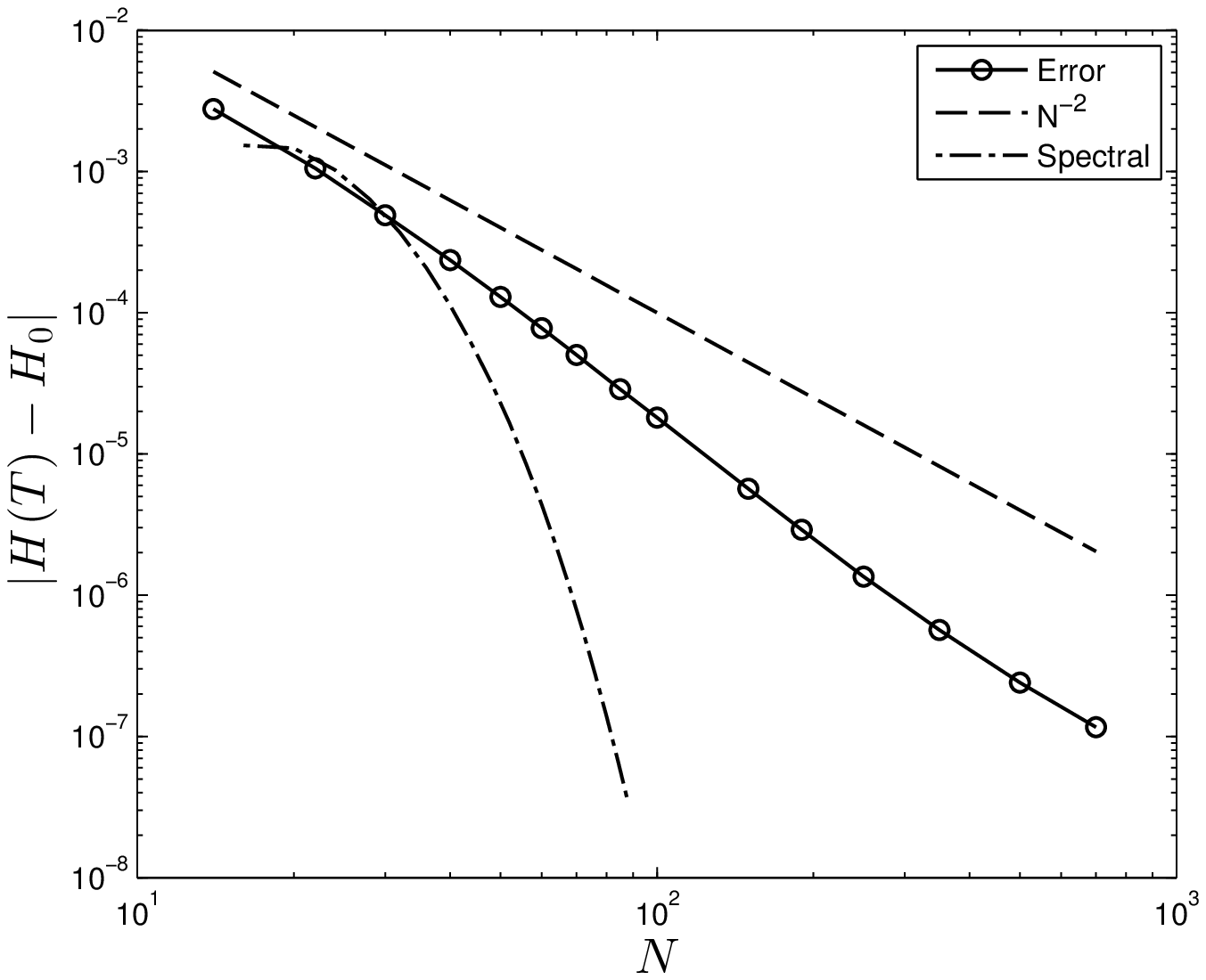}}
  \subfigure[Momentum $Q$]%
  {\includegraphics[width=0.49\textwidth]{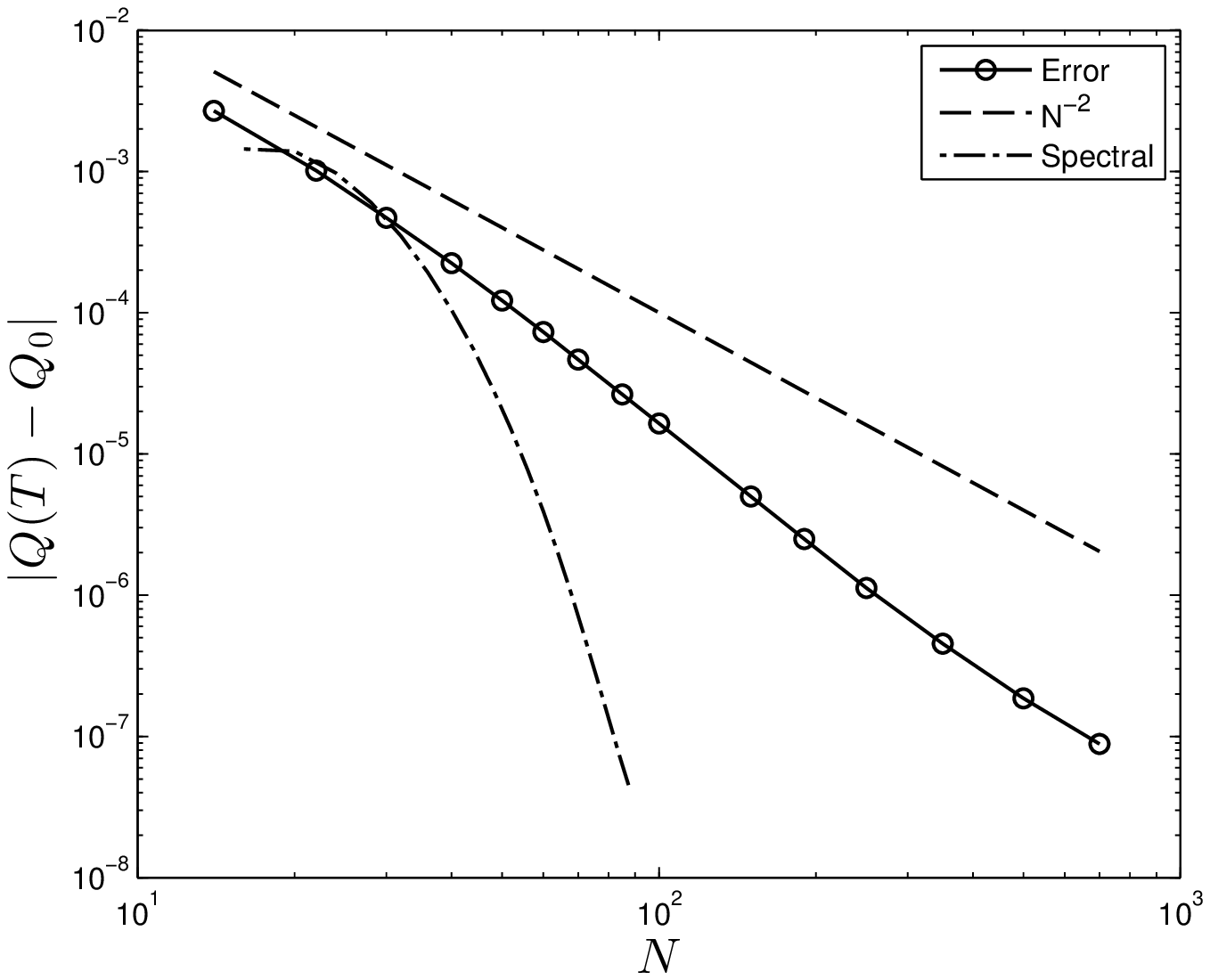}}
  \caption{\em Hamiltonian and generalized momentum conservation convergence computed using the finite volume and spectral methods under the mesh refinement. The conserved quantities are measured at the final simulation time.}
  \label{fig:HQ}
\end{figure}

The error is measured using the discrete $L_\infty$ norm for various successively refined discretizations. The result is shown on Figure~\ref{fig:conv}. As anticipated, the finite volume scheme (black solid line with circles) shows a fairly good second order convergence (with estimated slope $\approx 1.99$). During all the numerical tests, the mass conservation was satisfied with accuracy of the order $\approx 10^{-14}$. This impressive result is due to excellent local conservative properties of the finite volume method. We also investigate the numerical behavior of the scheme with respect to the less obvious invariants $\H$ and $Q$ defined in \eqref{eq:HQ}. These invariants can be computed exactly for solitary waves. However, we do not provide them to avoid cumbersome expressions. For the solitary wave with parameters given in Table~\ref{tab:params}, the generalized energy and momentum are given by the following expressions:
\begin{align*}
  \H_0\ =&\ \frac{21\sqrt{7}}{100}\ +\ \frac{7\sqrt{3}}{10}\log\frac{\sqrt{21}-1}{\sqrt{21}+1}\ 
  \approx\ 0.0178098463, \\
  Q_0\ =&\ \frac{62\sqrt{15}}{225}\ +\ \frac{2\sqrt{35}}{5}\log\frac{\sqrt{21}-1}{\sqrt{21}+1}\ 
  \approx\ 0.017548002.
\end{align*}
These values are used to measure the error on these quantities at the end of the simulation. Convergence of this error under the mesh refinement is shown on Figure~\ref{fig:HQ}. One can observe a slight super-convergence phenomenon of the finite volume scheme. This effect is due to the special nature of the solution we use to measure the convergence. This solution is only translated under the system dynamics. For more general initial conditions we expect a fair theoretical $2$\up{nd} order convergence for the finite volume scheme. As anticipated, the pseudo-spectral scheme shows the exponential error decay.

\subsection{Solitary wave interaction}

Solitary wave interactions are an important phenomenon in nonlinear dispersive waves which have been studied by numerical and analytical methods and results have been compared to experimental evidence. They also often serve as one of the most robust nonlinear benchmark test cases for numerical methods. We mention only a few works among the existing literature. For example, in \cite{Maxworthy1976, Renouard1985, CGHHS} solitary wave interactions were studied experimentally. The head-on collision of solitary waves was studied in the framework of full Euler equations in \cite{CGHHS, Chambarel2009}. Studies of solitary waves in various approximate models can be found in \cite{Li2004, Dougalis2004, ADM1, Dutykh2011e, Dutykh2010e}. To our knowledge, solitary wave collisions for the Serre equations were studied numerically for the first time in the PhD thesis of Seabra-Santos \cite{Seabra-Santos1985}. Finally, there are also a few studies devoted to simulations with full Euler equations \cite{Li2004, Fructus2005, CGHHS}.

\subsubsection{Head-on collision}

\begin{figure}
  \centering
  \includegraphics[width=0.90\textwidth]{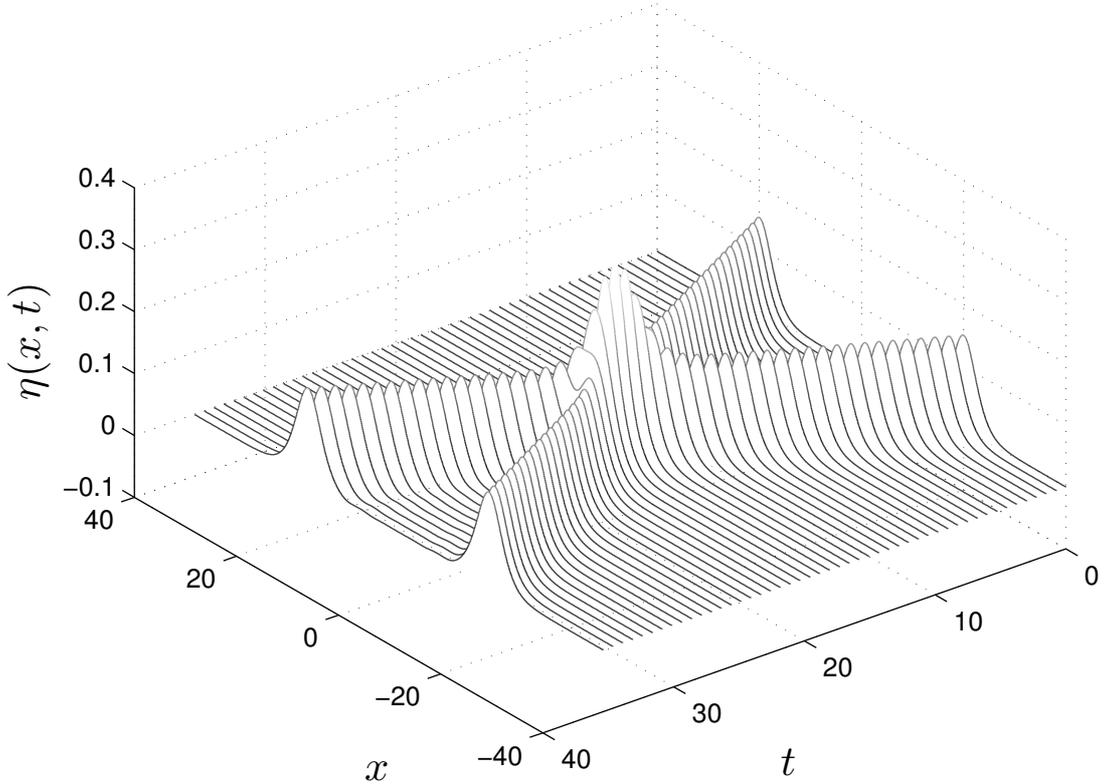}
  \caption{\em Head-on collision of two equal solitary waves simulated with the finite volume scheme.}
  \label{fig:headonfv}
\end{figure}

Consider the Serre equations posed in the domain $[-40, 40]$ with periodic boundary conditions. In the present section, we study the head-on collision (weak interaction) of two solitary waves of equal amplitude moving in opposite directions. Initially, two solitary waves of amplitude $a = 0.15$ are located at $x_0 = \pm 20$ (other parameters can be found in Table~\ref{tab:params}). The computational domain is divided into $N = 1000$ intervals (finite volumes in 1D) of the uniform length $\Delta x = 0.08$. The time step is chosen to be $\Delta t\approx 10^{-3}$. The process is simulated up to time $T = 36$. The numerical results are presented in Figure~\ref{fig:headonfv}. As expected, the solitary waves collide quasi-elastically and continue to propagate in opposite directions after the interaction. The value of importance is the maximum amplitude during the interaction process, sometimes referred to as the run-up. Usually, it is larger than the sum of the amplitudes of the two initial solitary waves. In this case, we obtain a run-up of $0.3130 > 2a = 0.3$.

In order to validate the finite volume simulation, we performed the same computation with the pseudo-spectral method presented briefly in Section~\ref{sec:spectral}. We used a fine grid of $1024$ nodes and adaptive time stepping. The overall interaction process is visually identical to the finite volume result shown in Figure~\ref{fig:headonfv}. The run-up value according to the spectral method is $0.3127439$ showing again the accuracy of our simulation. The small inelasticity is evident from the small dispersive wave train emerging after the interaction (for an example in a slightly different setting described below, see Figure~\ref{fig:ho205}, as first found numerically and experimentally by Seabra-Santos \cite{Seabra-Santos1985}.

\subsubsection{Overtaking collision}

\begin{figure}
  \centering
  \includegraphics[width=0.90\textwidth]{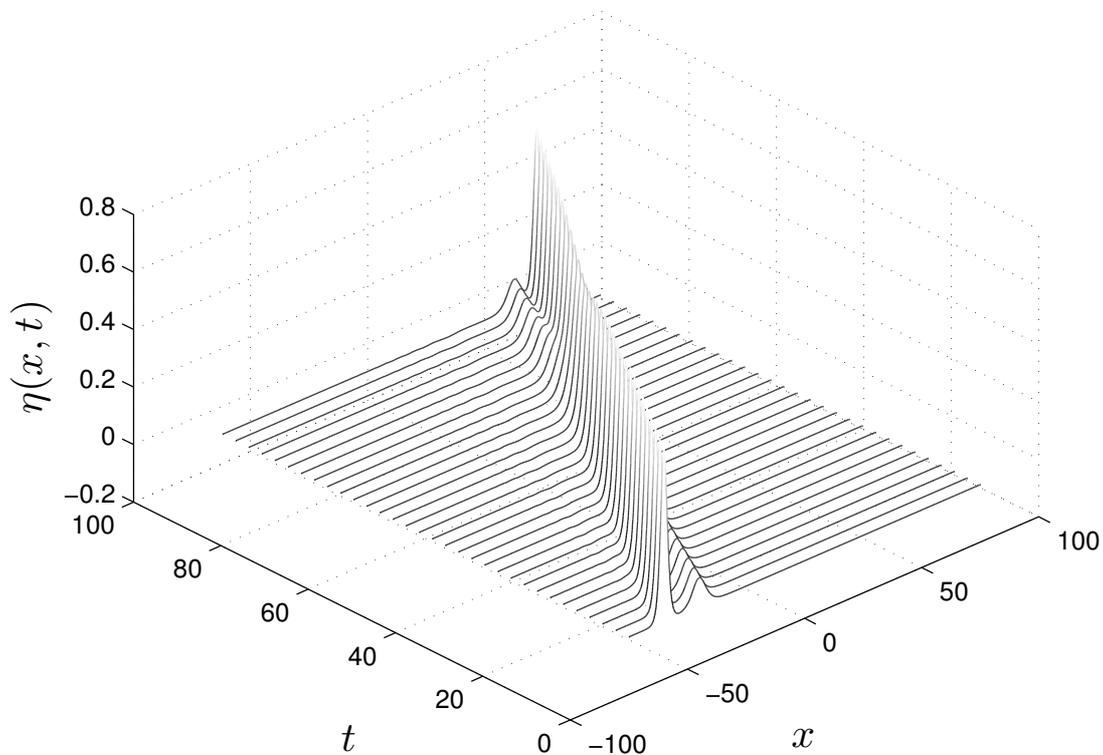}
  \caption{\em Overtaking (or following) collision of two solitary waves simulated with the finite volume scheme.}
  \label{fig:followcol}
\end{figure}

\begin{figure}
  \centering
  \includegraphics[width=0.75\textwidth]{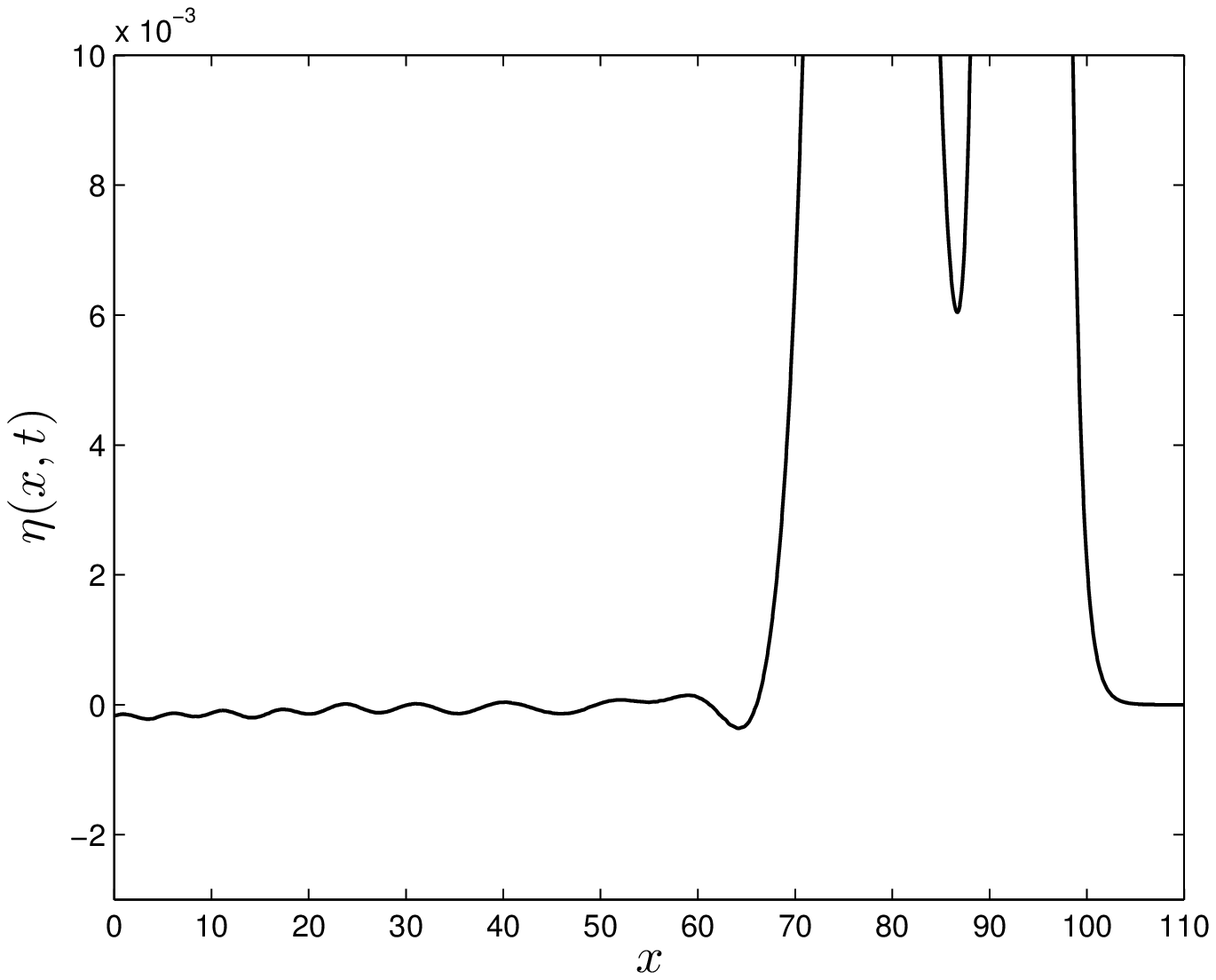}
  \caption{\em Dispersive tail after overtaking collision of two solitary waves (strong interaction) at $T = 120.0$.}
  \label{fig:hoT120}
\end{figure}

A second type of solitary wave interaction is the \emph{overtaking collision} (or \emph{strong interaction}) of two solitary waves of different amplitudes moving in the same direction. Sometimes this situation is also referred to as the following collision or strong interaction. For this case we consider a physical domain $[-75, 75]$ divided into $N = 1000$ equal control volumes.The initial data consists of two separated solitary waves of different amplitudes moving in the same direction. The solitary wave with larger amplitude moves faster and will overtake the smaller wave. This situation was simulated with the finite volume scheme and the numerical results are presented in Figure~\ref{fig:followcol}. The parameters used in this simulation are given in Table~\ref{tab:params2}. The strong interaction is also inelastic with a small dispersive tail emerging after the over-taking (see Figure~\ref{fig:hoT120} for a zoom).

\begin{table}
  \centering
	\begin{tabular}{l|c}
	  \hline\hline
	  Undisturbed water depth: $d$ & 1 \\
	  Gravity acceleration: $g$ & 1 \\
	  Large solitary wave amplitude: $a_1$ & 0.6 \\
	  Initial position: $x_1$ & -60 \\
	  Small solitary wave amplitude: $a_2$ & 0.1 \\
	  Initial position: $x_2$ & -45 \\
	  Final simulation time: $T$ & 96 \\
	  Free parameter: $\beta$ & 1/3 \\
	  \hline\hline
	\end{tabular}
  \caption{\em Values of various parameters used to simulate the overtaking collision.}
  \label{tab:params2}
\end{table}

\begin{figure}
  \centering
  \subfigure[$t = 18.5$ s]%
  {\includegraphics[width=0.49\textwidth]{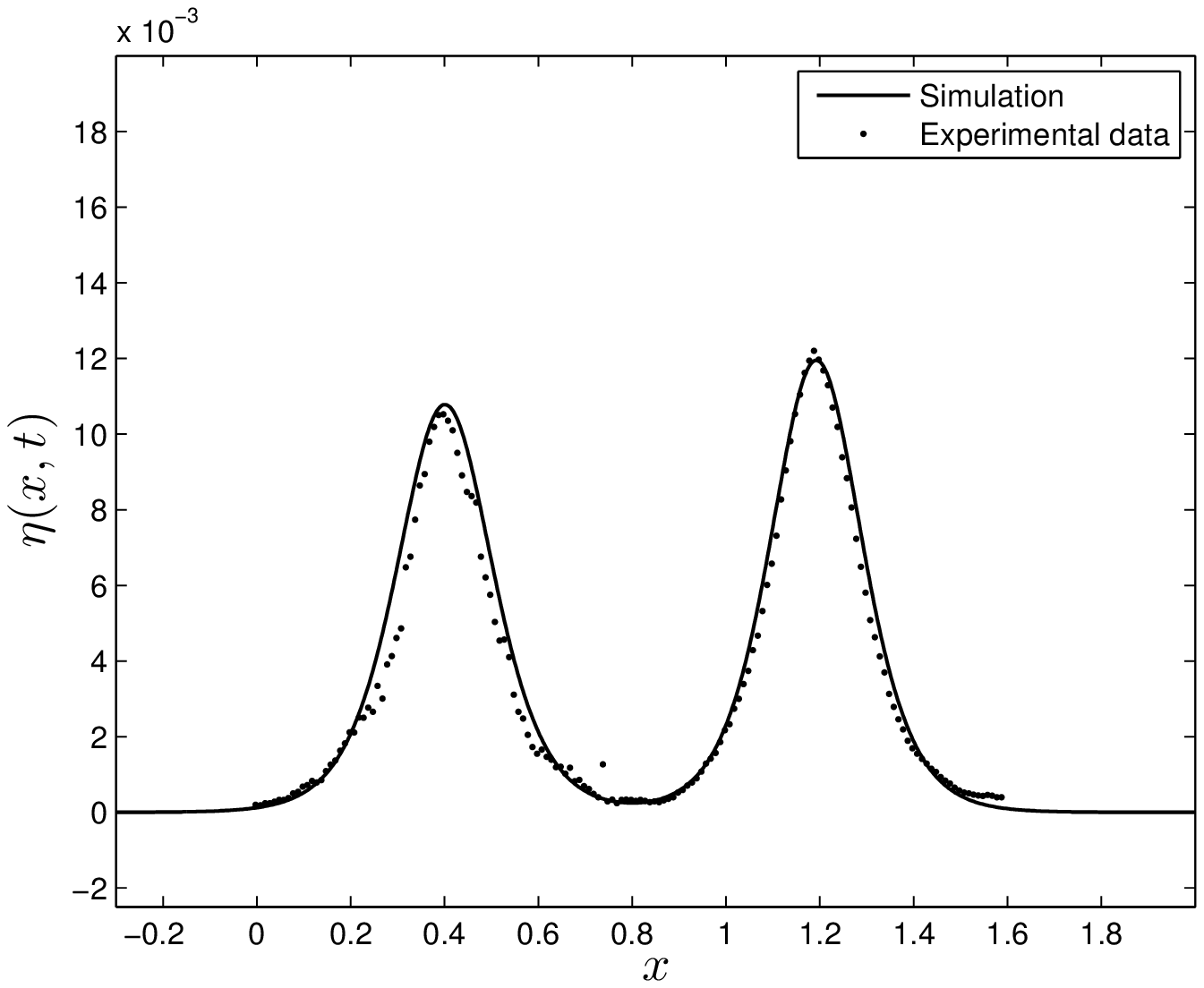}}
  \subfigure[$t = 18.6$ s]%
  {\includegraphics[width=0.49\textwidth]{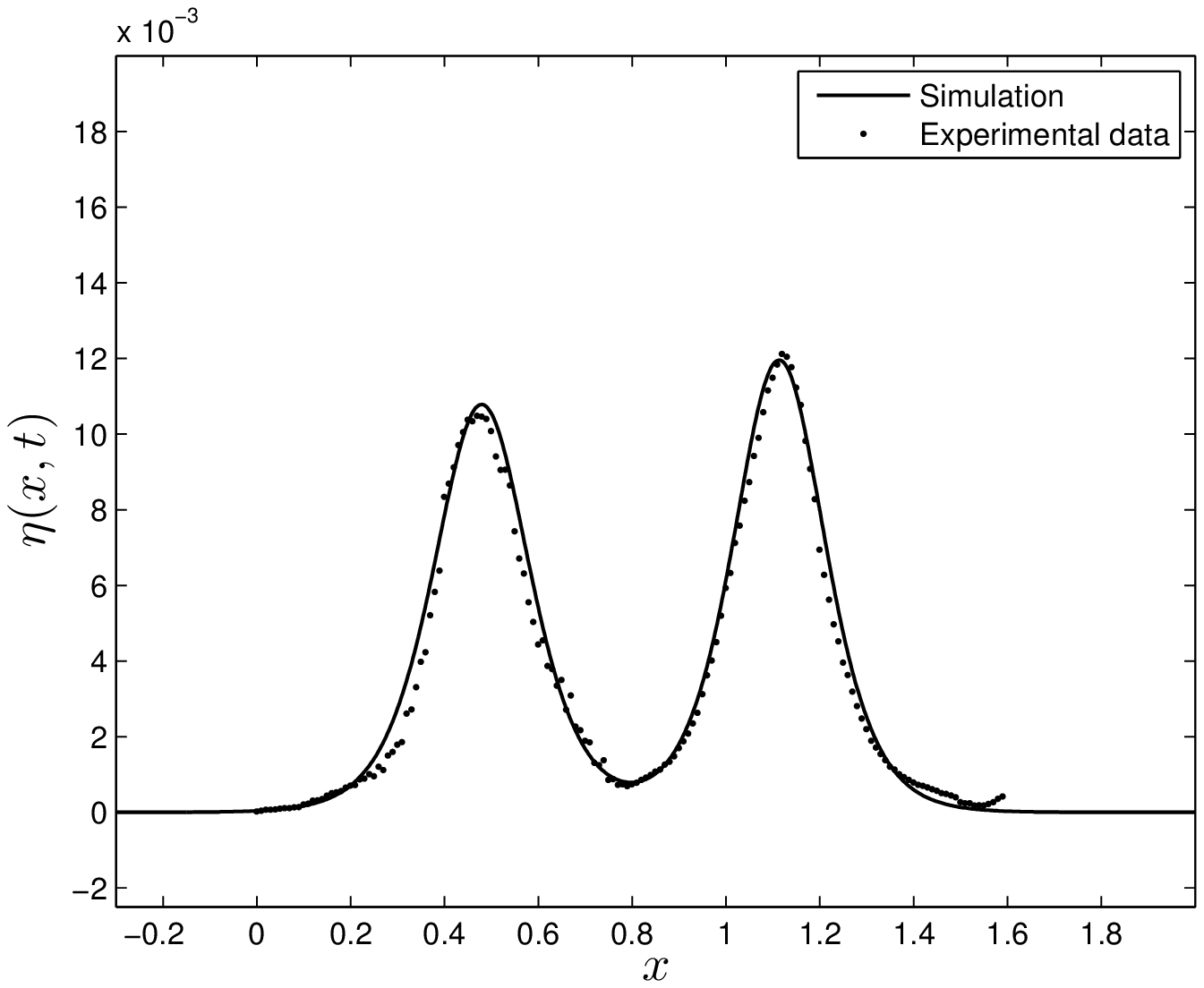}}
  \caption{\em Head-on collision of two solitary waves of different amplitudes. Comparison with experimental data \cite{CGHHS}.}
  \label{fig:ho1}
\end{figure}

\begin{figure}
  \centering
  \subfigure[$t = 18.7$ s]%
  {\includegraphics[width=0.49\textwidth]{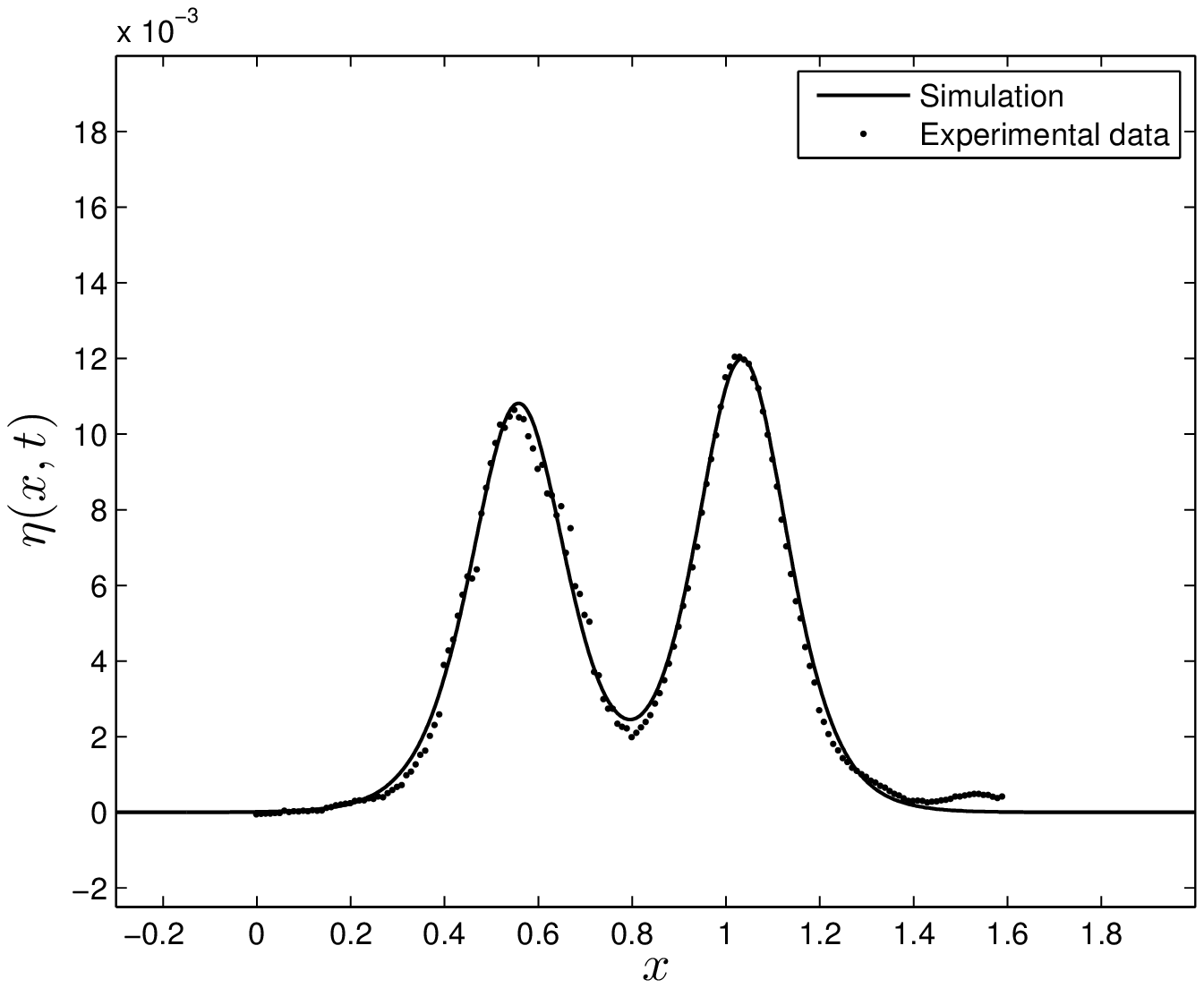}}
  \subfigure[$t = 18.8$ s]%
  {\includegraphics[width=0.49\textwidth]{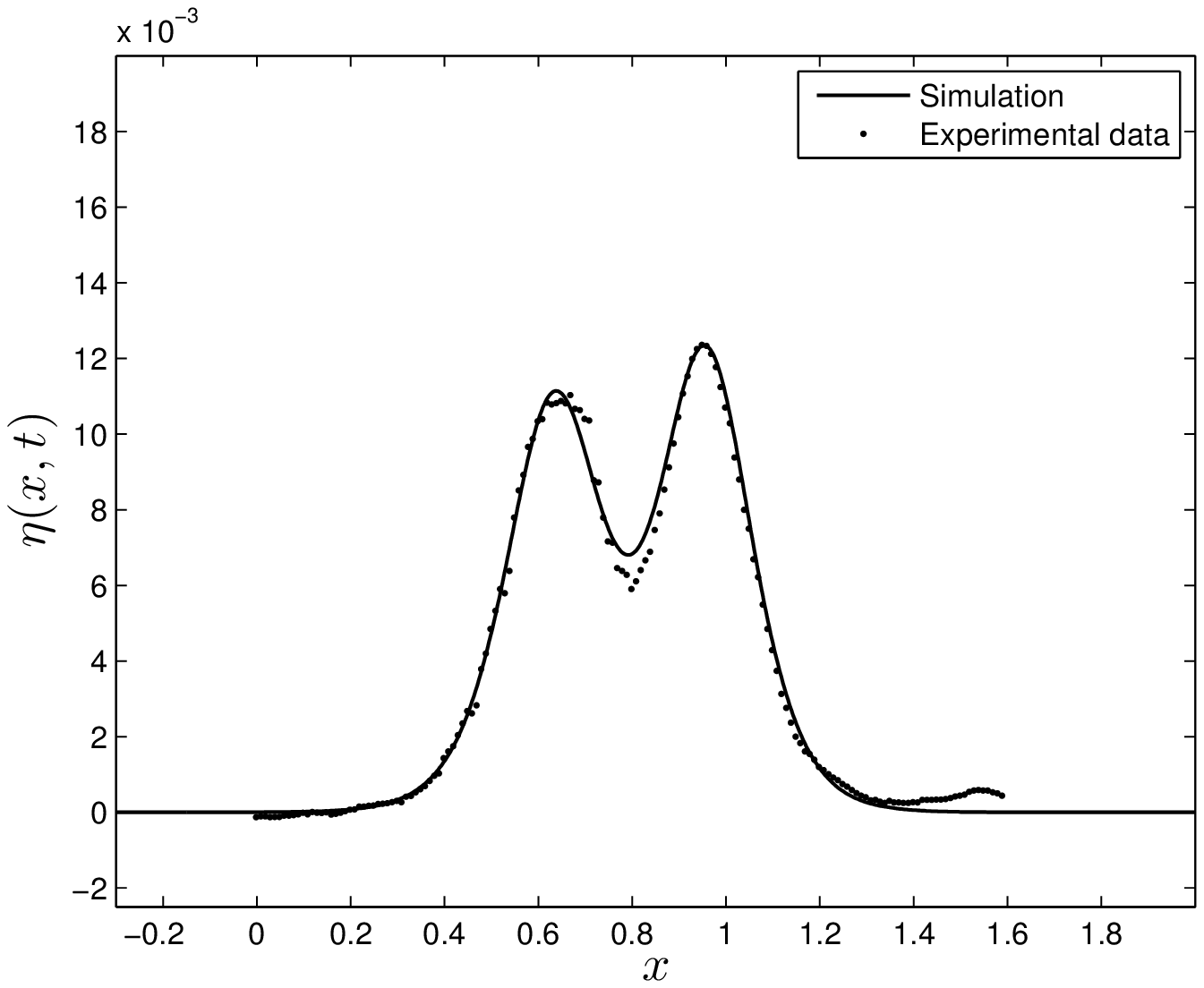}}
  \caption{\em Head-on collision of two solitary waves of different amplitudes. Comparison with experimental data \cite{CGHHS}.}
\end{figure}

\begin{figure}
  \centering
  \subfigure[$t = 18.92$ s]%
  {\includegraphics[width=0.49\textwidth]{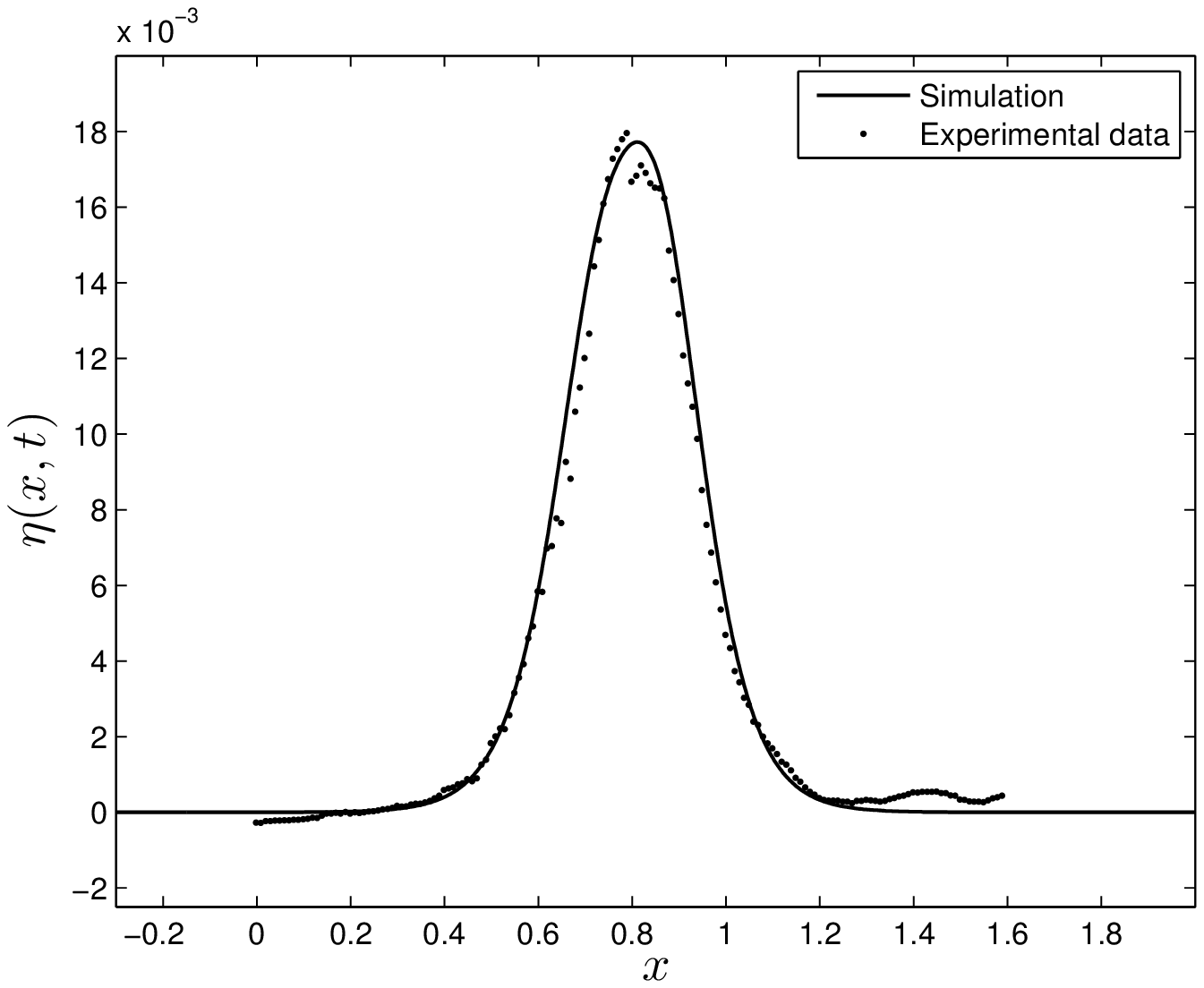}}
  \subfigure[$t = 19.0$ s]%
  {\includegraphics[width=0.49\textwidth]{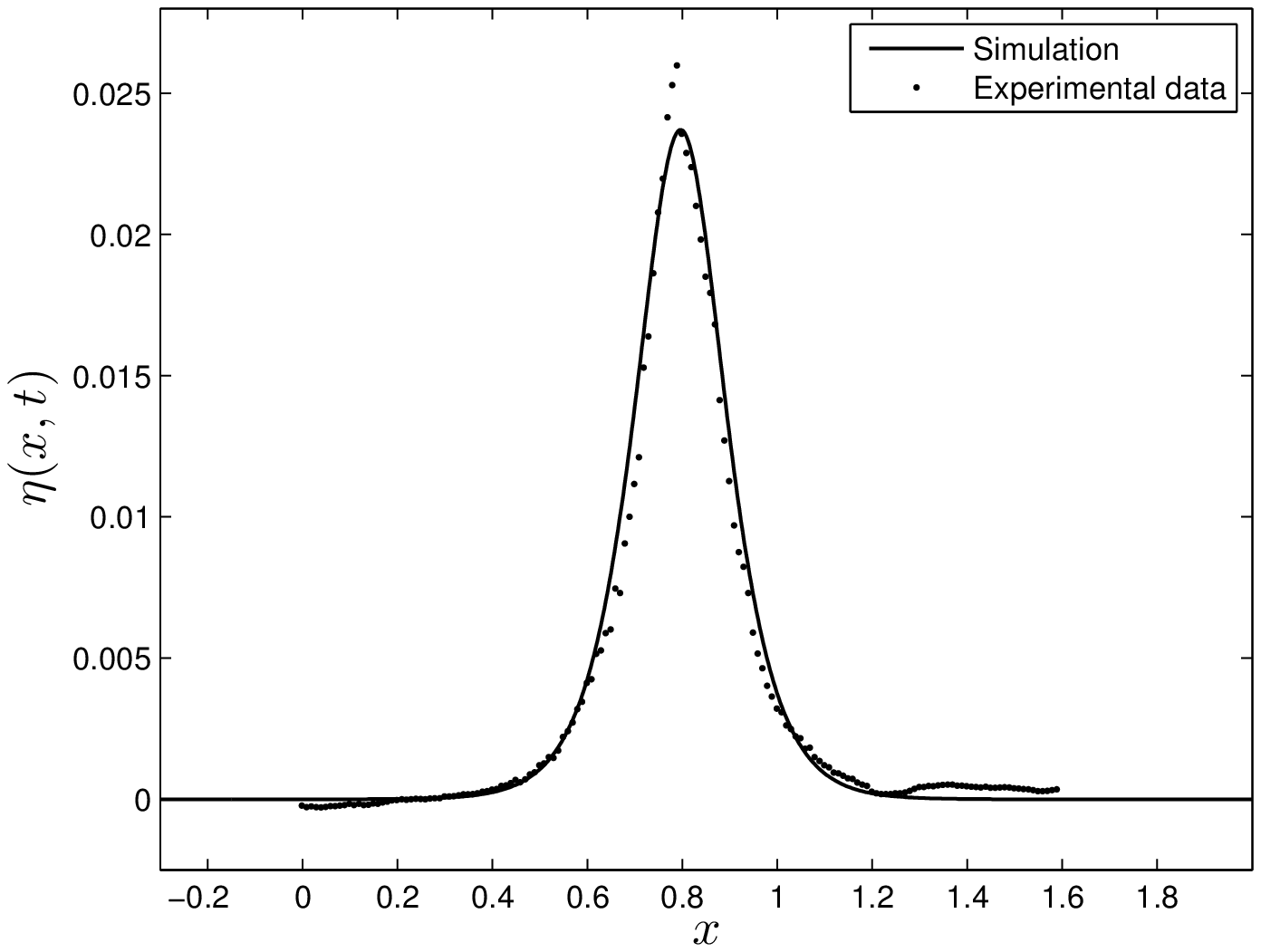}}
  \caption{\em Head-on collision of two solitary waves of different amplitudes. Comparison with experimental data \cite{CGHHS}. Note the difference in vertical scales on the left and right images.}
\end{figure}

\begin{figure}
  \centering
  \subfigure[$t = 19.05$ s]%
  {\includegraphics[width=0.49\textwidth]{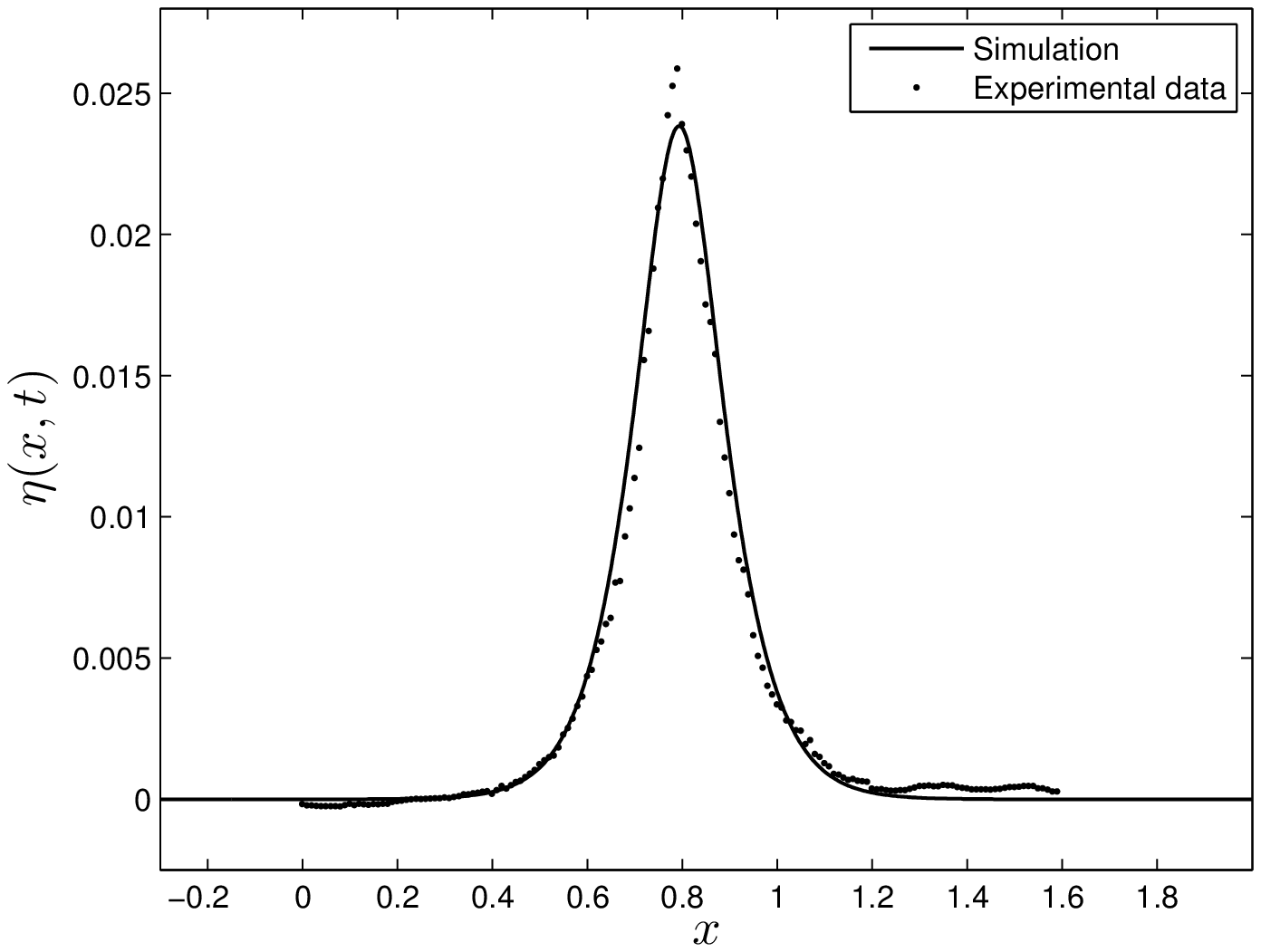}}
  \subfigure[$t = 19.1$ s]%
  {\includegraphics[width=0.49\textwidth]{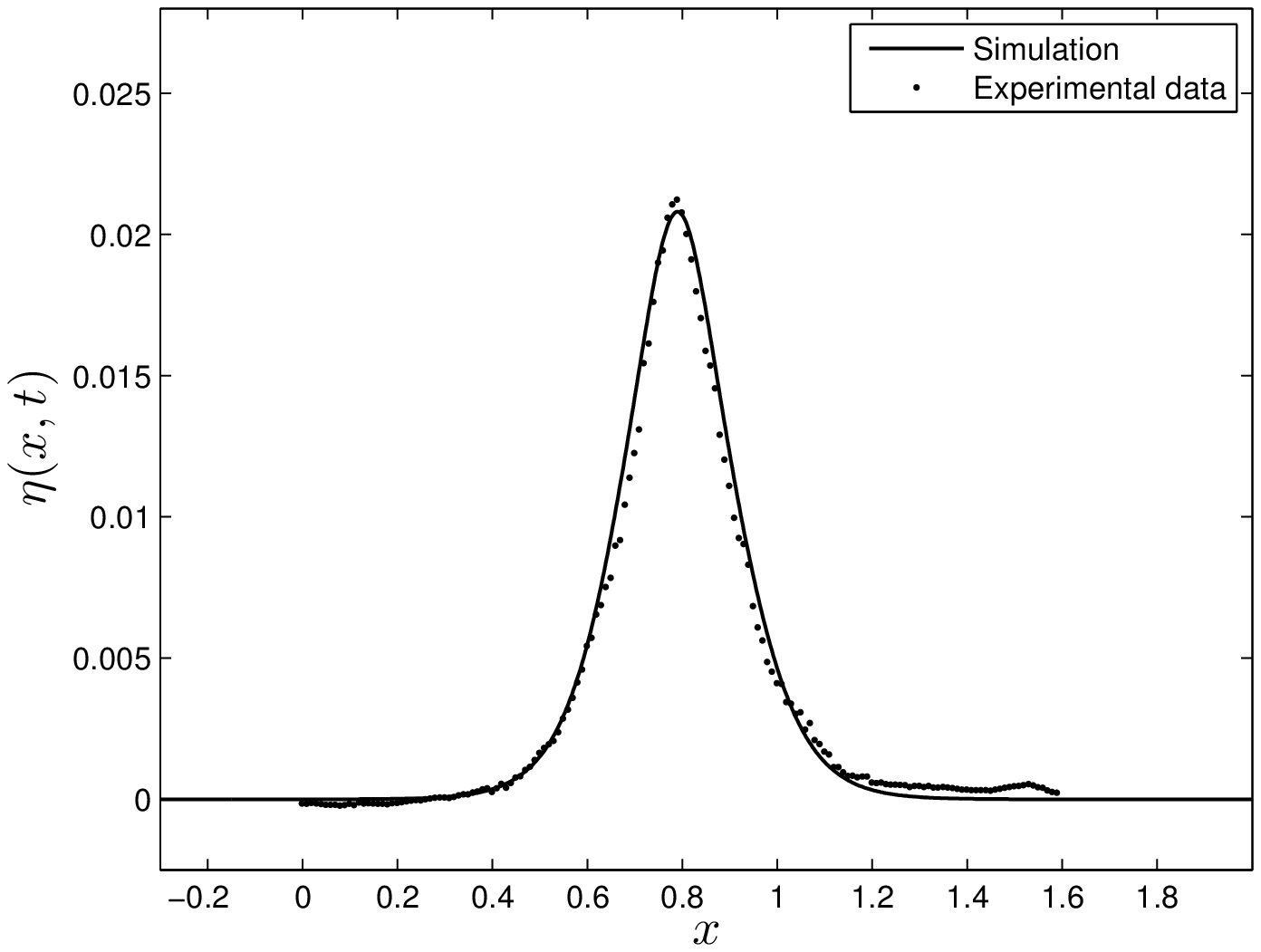}}
  \caption{\em Head-on collision of two solitary waves of different amplitudes. Comparison with experimental data \cite{CGHHS}.}
\end{figure}

\begin{figure}
  \centering
  \subfigure[$t = 19.15$ s]%
  {\includegraphics[width=0.49\textwidth]{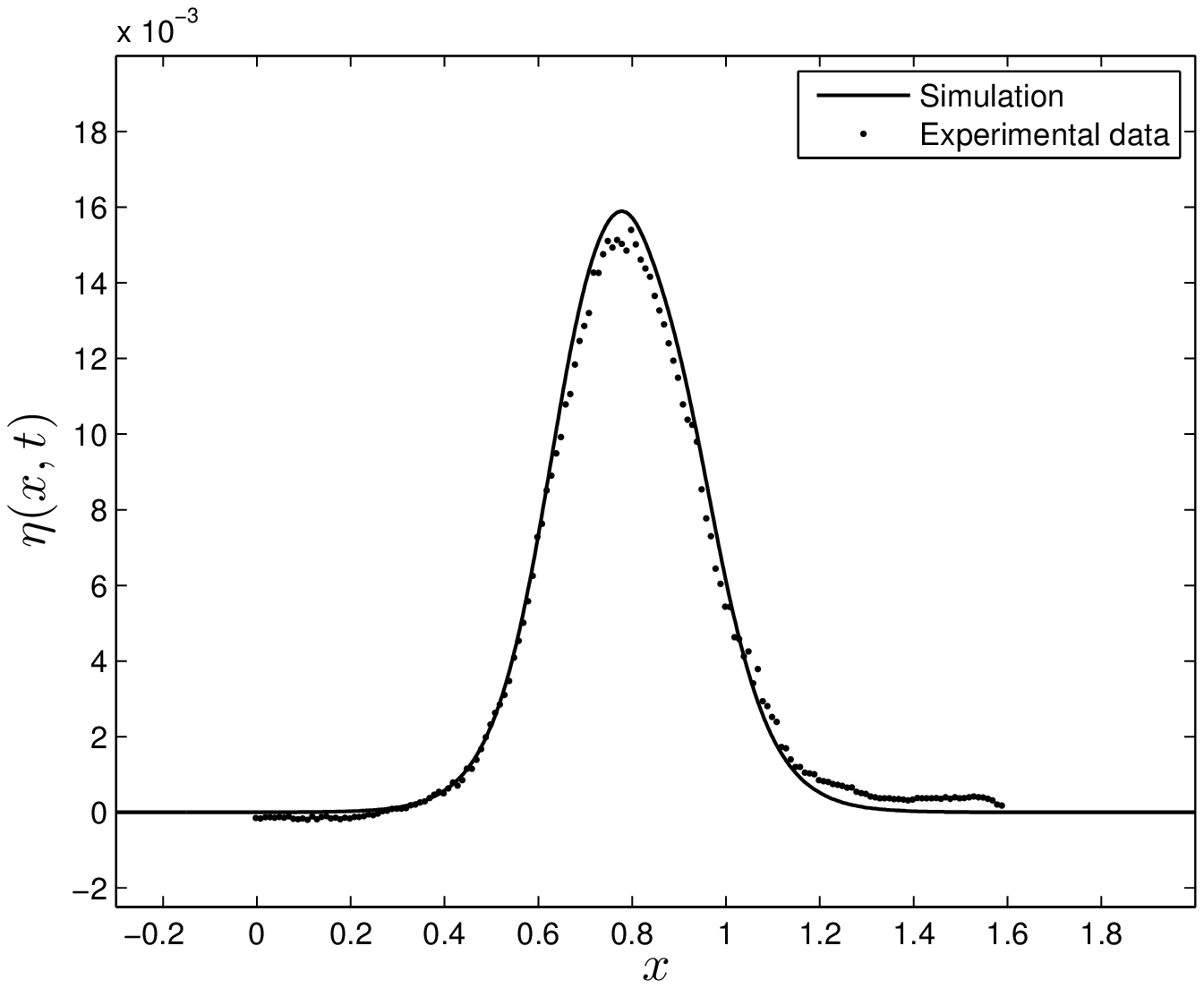}}
  \subfigure[$t = 19.19$ s]%
  {\includegraphics[width=0.49\textwidth]{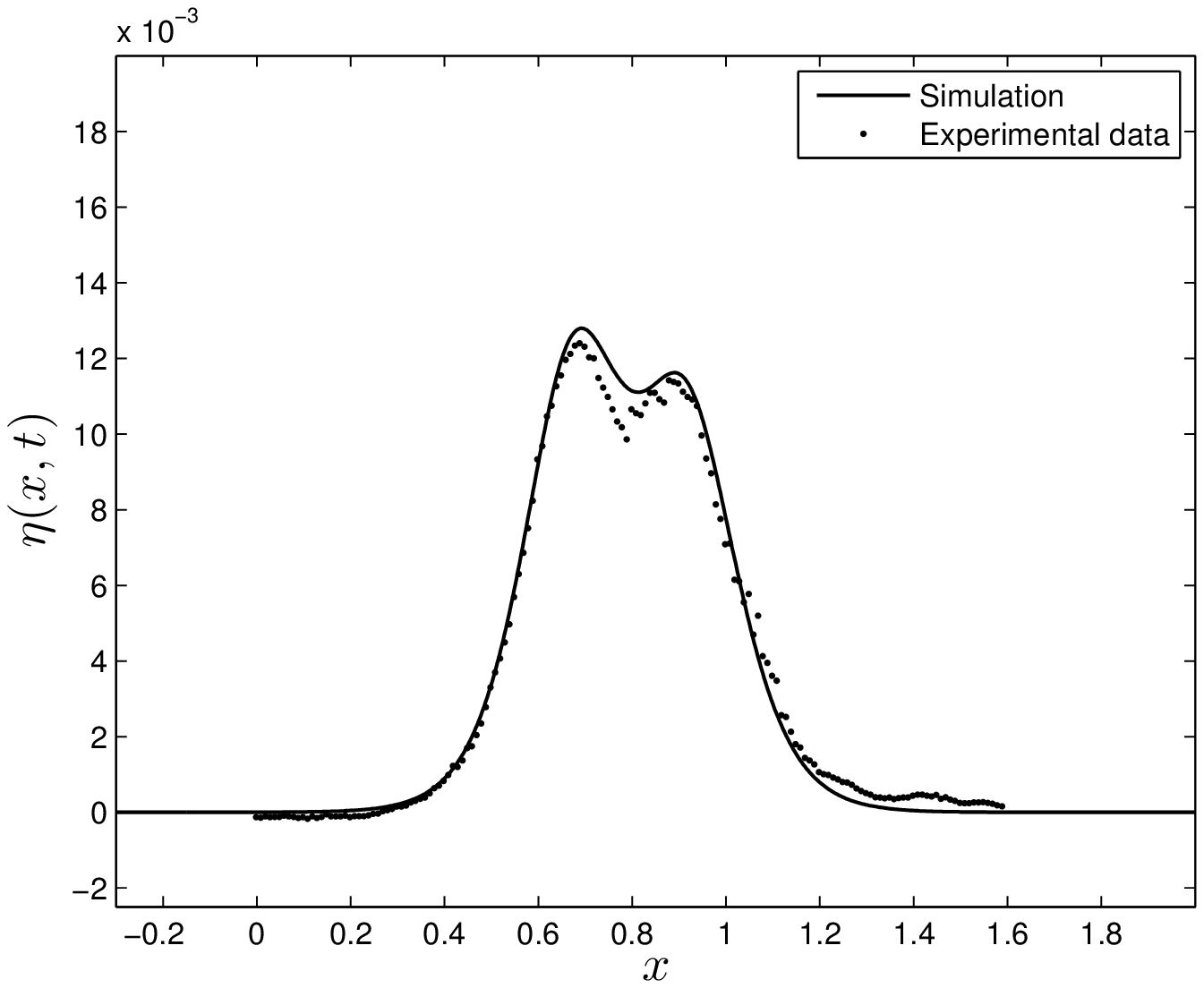}}
  \caption{\em Head-on collision of two solitary waves of different amplitudes. Comparison with experimental data \cite{CGHHS}.}
\end{figure}

\begin{figure}
  \centering
  \subfigure[$t = 19.33$ s]%
  {\includegraphics[width=0.49\textwidth]{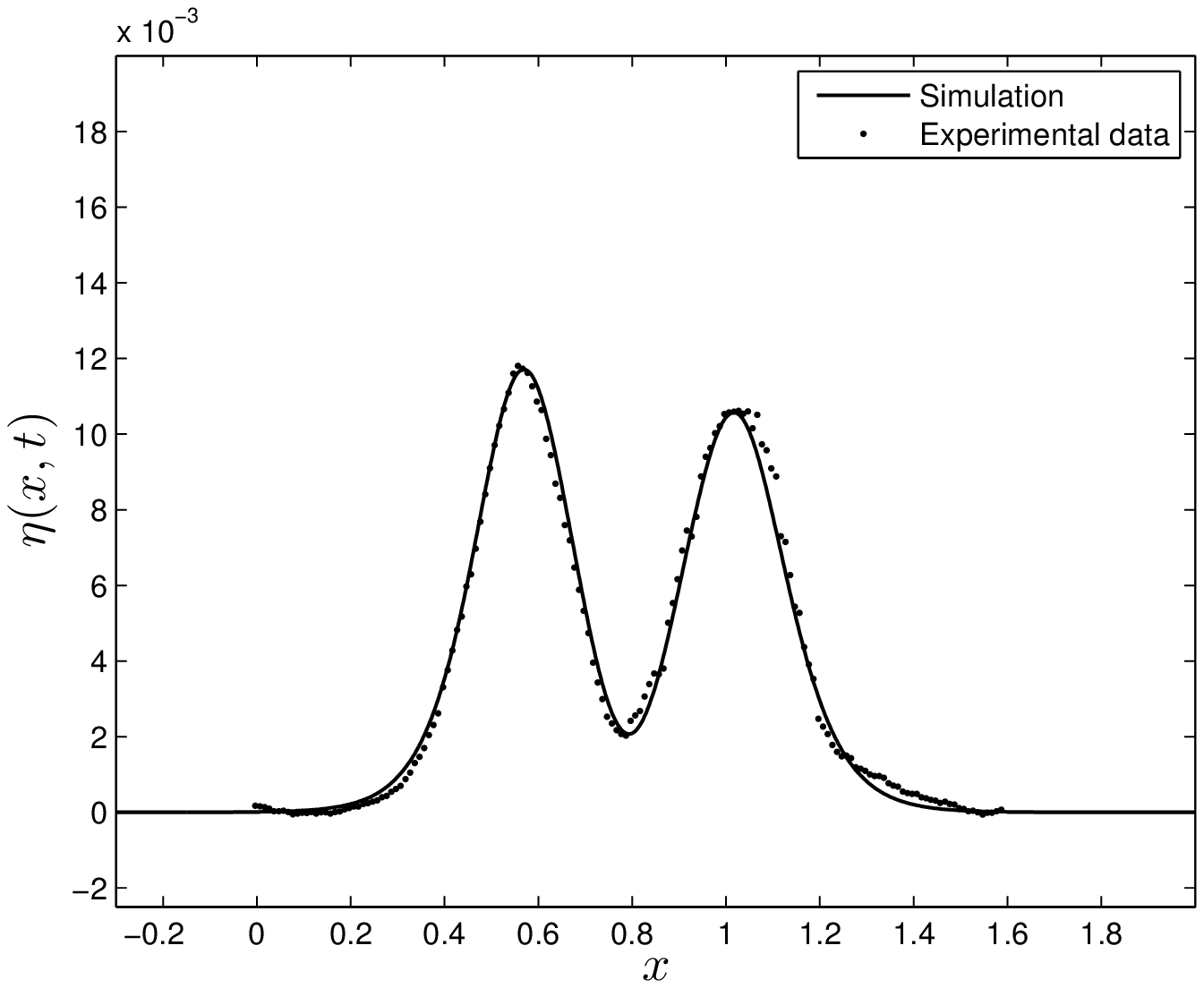}}
  \subfigure[$t = 19.5$ s]%
  {\includegraphics[width=0.49\textwidth]{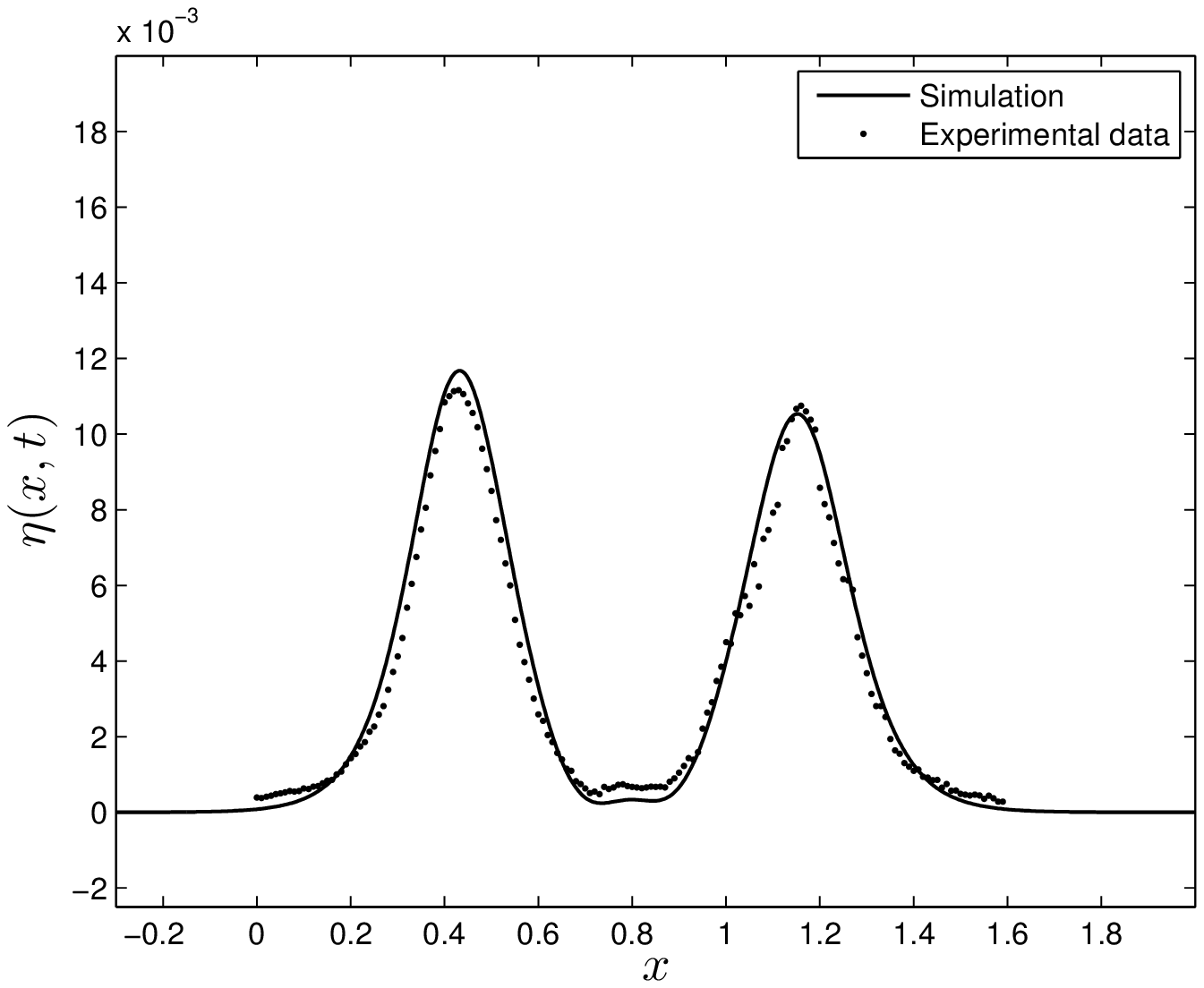}}
  \caption{\em Head-on collision of two solitary waves of different amplitudes. Comparison with experimental data \cite{CGHHS}.}
\end{figure}

\begin{figure}
  \centering
  \subfigure[$t = 19.85$ s]%
  {\includegraphics[width=0.49\textwidth]{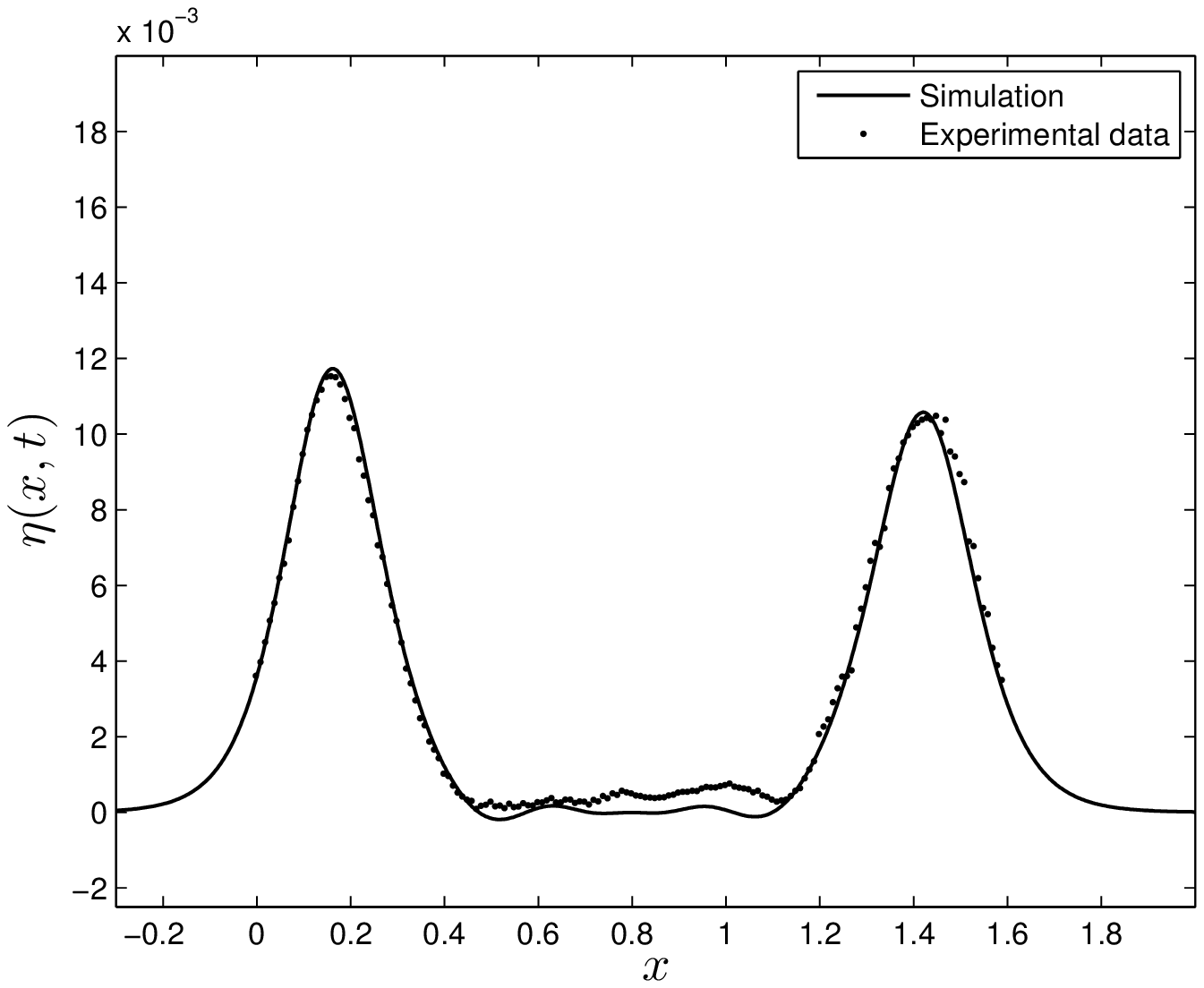}}
  \subfigure[$t = 20.0$ s]%
  {\includegraphics[width=0.49\textwidth]{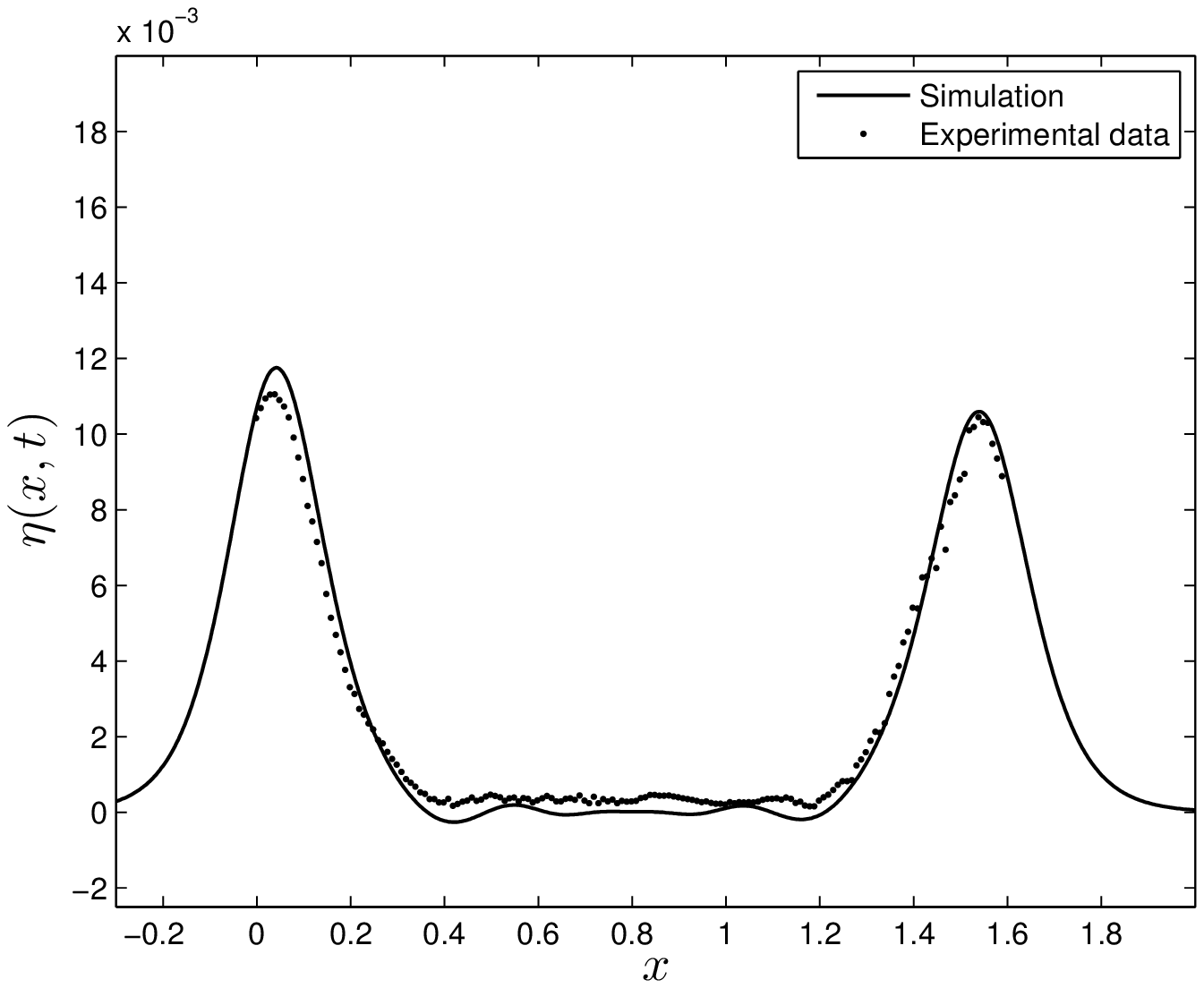}}
  \caption{\em Head-on collision of two solitary waves of different amplitudes. Comparison with experimental data \cite{CGHHS}.}
\end{figure}

\subsection{Experimental validation}

In this Section we present a comparison between the classical Serre model solved with our finite volume scheme and one head-on collision experiment from \cite{CGHHS}. This specific experiment was already considered in the context of Boussinesq-type systems \cite{Dutykh2011e}.

\begin{figure}
  \centering
  \includegraphics[width=0.75\textwidth]{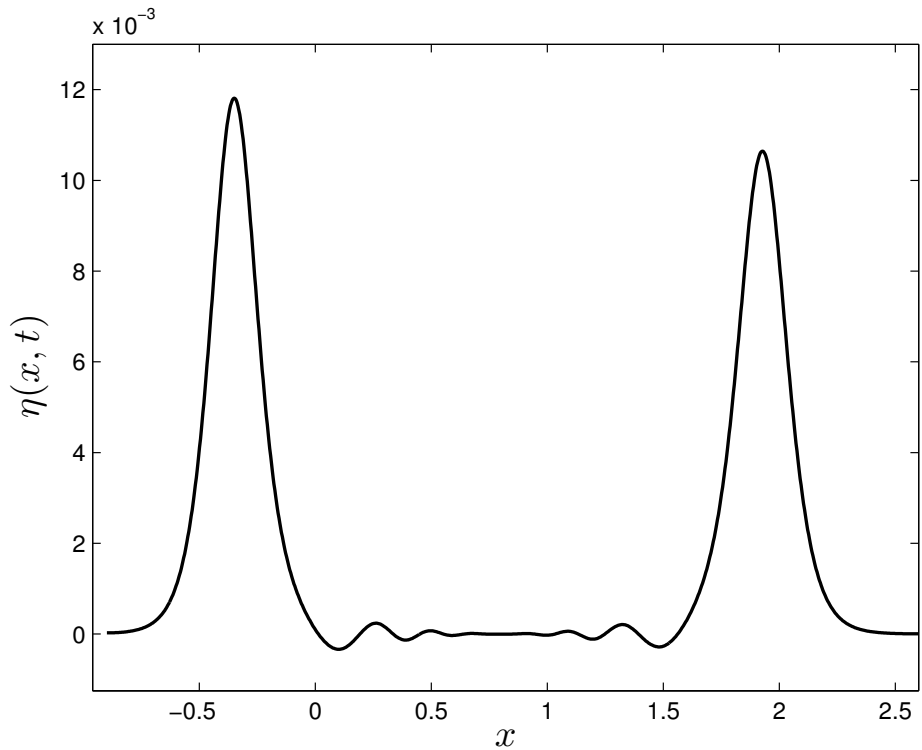}
  \caption{\em Dispersive tail after head-on collision of two solitary waves (weak interaction). Small wavelets between two solitary waves clearly indicate that the collision is inelastic.}
  \label{fig:ho205}
\end{figure}

We simulate a portion of the wave tank $[-0.9, 2.7]$ (divided into $N = 1000$ equal control volumes) where the interaction process takes place. The initial data consists of two solitary waves (of different amplitudes in this case) moving in opposite directions. The exact parameters are given in Table~\ref{tab:params3}. Simulation snapshots are presented in Figures \ref{fig:ho1}--\ref{fig:ho205}. The general agreement is very good, validating the Serre equations in water wave theory, along with our numerical developments. Figure~\ref{fig:ho205} shows visible dispersive oscillations after the interaction process, numerical evidence of the inelastic character of solitary waves interactions in the framework of the Serre equations.

\begin{table}
  \centering
	\begin{tabular}{l|c}
	  \hline\hline
	  Undisturbed water depth: $d$ [{\sf cm}] & 5 \\
	  Gravity acceleration: $g$ [$\mathsf{m}\,\mathsf{s}^{-2}$] & 9.81 \\
	  Right-going SW amplitude: $a_1$ [{\sf cm}] & 1.077 \\
	  Initial position of the SW-1: $x_1$ [{\sf m}] & 0.247 \\
	  Left-going SW amplitude: $a_1$ [{\sf cm}] & 1.195 \\
	  Initial position of the SW-2: $x_2$ [{\sf m}] & 1.348 \\
	  Final simulation time: $T$ [{\sf s}] & 20.5 \\
	  \hline\hline
	\end{tabular}
  \caption{\em Values of various parameters used to simulate the head-on collision.}
  \label{tab:params3}
\end{table}

\section{Conclusions}\label{sec:concl}

The current study is devoted to the Serre equations stemming from water wave modeling \cite{Serre1953, Barthelemy2004, Dias2010}. First, we presented a derivation of this model using a relaxed variational principle \cite{Clamond2009}. We then described an Implicit-Explicit finite volume scheme to discretize the equations. The overall theoretical accuracy of the discretization scheme is of second-order. This conclusion is confirmed by comparisons with an exact solitary wave solution. The energy conservation properties of our scheme are also discussed and quantified. In order to validate further our numerical scheme, we present a Fourier-type pseudo-spectral method. Both numerical methods are compared on solitary wave interaction problems. The proposed discretization procedure was successfully validated with several numerical tests along with experimental data. In contrast with the highly accurate spectral method, the finite volume method has the advantage of being robust and generalizable to realistic complex situations with variable bathymetry, very steep fronts, dry areas, etc. The present study should be considered as the first step to further generalisations to 2D cartesian meshes \cite{Mingham1998, Causon2000, Vignoli2008}.

\subsection*{Acknowledgments}
\addcontentsline{toc}{section}{Acknowledgments}

D.~\textsc{Dutykh} acknowledges the support from French ``Agence Nationale de la Recherche'', project ``MathOc\'ean'' (Grant ANR-08-BLAN-0301-01) along with the support from ERC under the research project ERC-2011-AdG 290562-MULTIWAVE. P.~\textsc{Milewski} acknowledges the support of the University of Savoie during his visits in 2011.

\addcontentsline{toc}{section}{References}
\bibliographystyle{abbrv}
\bibliography{biblio}

\end{document}